\documentstyle[manuscript,eqsecnum,aps]{revtex}
\tighten
\begin{document}
\gdef\journal#1, #2, #3, 1#4#5#6{               
    {\sl #1~}{\bf #2}, #3, (1#4#5#6)}           
\draft
\preprint{UCSBTH-94-04}
\date{\today}

\title{Spacetime Information}

\author{James B. Hartle\thanks{hartle@cosmic.physics.ucsb.edu}}
\vskip .13 in
\address{Department of Physics,\\
 University of California,
Santa Barbara, CA
93106-9530}
\address{and}
\address{Isaac Newton Institute for Mathematical Sciences,\\
 University of
Cambridge, Cambridge CB3 0EH, UK}

\maketitle

\begin{abstract}

In usual quantum theory, the information available about a quantum
system is defined in terms of the density matrix describing it on a
spacelike surface.  This definition must be generalized for extensions
of quantum theory which neither require, nor always permit, a notion of
state on a spacelike surface.  In particular, it must be generalized for
the generalized quantum theories appropriate when spacetime geometry
fluctuates quantum mechanically or when geometry is fixed but not foliable by
spacelike surfaces.  This paper introduces a four-dimensional notion of
the information available about a quantum system's boundary conditions
in the various sets of
decohering, coarse-grained histories it may display.  This spacetime
notion of information coincides with the familiar one when quantum theory
{\it is} formulable in terms of states on spacelike surfaces but generalizes
this notion when it cannot be so formulated.  The idea of spacetime
information is applied in several contexts: When spacetime geometry is
fixed the information available through alternatives restricted to a
fixed spacetime region is defined. The information
available through histories of alternatives of general operators is
compared to that obtained from the more limited coarse-grainings of
sum-over-histories quantum mechanics that refer only to co\"ordinates.
The definition of information is considered in generalized quantum
theories. We consider as specific examples time-neutral quantum
mechanics with
 initial and final conditions, quantum
theories with non-unitary evolution, and the generalized quantum
frameworks appropriate for quantum spacetime.  In such theories complete
information about a quantum system is not necessarily available on any
spacelike surface but must be searched for throughout spacetime.  The
information loss commonly associated with the ``evolution of pure states
into mixed states'' in black hole evaporation is thus not in conflict
with the principles of generalized quantum mechanics.

\end{abstract}
\vfill
\pacs{PACS number: 03.65.Bz, 04.60-m, 04.70.Dy, 98.80.Hw}

\setcounter{footnote}{0}
\section[]{Introduction}
\label{sec:I}

In the usual quantum theory of a system of matter fields in a fixed
background spacetime, the state of the fields on a spacelike Cauchy
surface is as complete a description of the system as it is possible to
give.  When the state is specified, the missing
information is zero.  If only probabilities $\pi_i$
 for the state to be one of a
set of states $\{|\psi_i(\sigma)\rangle\}$
 on a Cauchy surface $\sigma$ are specified, then the
missing information is greater.  The system may then be described by a
density matrix $\rho(\sigma)$

\begin{equation}
\rho(\sigma)= \Sigma_i |\psi_i(\sigma)\rangle \pi_i \langle\psi_i
(\sigma)|\ ,
\label{oneone}
\end{equation}
 and the missing information is
\begin{equation}
S(\sigma)= -\Sigma_i \pi_i\log \pi_i
= -Tr\bigl[\rho(\sigma) \log \rho(\sigma)\bigr]\ .
\label{onetwo}
\end{equation}
The unitary evolution of $\rho(\sigma)$ through a foliating family of
Cauchy surfaces ensures that $S(\sigma)$ defined by (\ref{onetwo}) is
independent of $\sigma$.  Complete information about a system is
obtainable on any Cauchy surface in the foliating family and that
information is the same on one surface as on any other.

However, when quantum fluctuations of spacetime geometry are taken into
account, as in any quantum theory of gravity, it is difficult to
formulate quantum theory in terms of states on spacelike surfaces.  This,
not least, because there is no fixed geometry to give a meaning to
``spacelike''.\footnote{For lucid reviews of the difficulties see
\cite{Kuc92,Ish92,Ish93,Unrpp91}.}  Similar difficulties exist when
spacetime geometry is fixed but not foliable by spacelike surfaces, as in
spacetimes with closed timelike curves ({\it e.g.,\ }
\cite{Har94}). A possible
approach to such situations is to generalize the quantum framework for
prediction so that is in fully spacetime form and does not require a
notion of ``state on a spacelike surface''.\footnote{As in \cite{Harpp}
where references to the earlier literature may be found.} How does one
discuss information when quantum mechanics
is in spacetime form and does not necessarily have a notion of state
on a spacelike surface
with which to define (\ref{onetwo})? This paper proposes an answer to
this question.

In a quantum theory fully in spacetime form it is appropriate to
take a four-dimensional,
 spacetime approach to the definition of information.  This is the
guiding principle of this paper. Applying ideas of
M.~Gell-Mann and the author \cite{GHup}, we implement this principle
in Sections II--IV in the usual formulation of the quantum mechanics of
a closed system. In Sections V--VIII we consider spacetime information
in generalized quantum theories \cite{Har91a,Harpp,Ish94}.

In the usual formulation of quantum mechanics of a closed system
 with a fixed background spacetime, probabilities for decohering sets of
histories are determined from an  initial
condition represented by a Heisenberg-picture density matrix $\rho$.
In Section II we define the information available about this initial
condition in any set of decoherent histories.  The minimum missing information
among all such sets defines the complete information available about the
system.  We show that in the usual formulation
 this spacetime notion of complete information
coincides with (\ref{onetwo}) and is available on every spacelike surface.

The information available in a spacetime region may be defined by
considering sets of histories representing alternatives that are
restricted to that region. This is discussed in Section III. A spacelike
surface is one kind of spacetime region and the information available on
it coincides with the usual definition (\ref{onetwo}).

In Section IV we compare the information available in a  quantum theory
that allows alternative values of all Hermitian operators (co\"ordinates,
momenta, etc.) with that available in a sum-over-histories formulation
of quantum mechanics that allows only alternatives defined by
paths in a set of generalized co\"ordinates.  Infinitely less
information is available on one spacelike surface in sum-over-histories
quantum mechanics than can be obtained
 by utilizing all possible observables. However,
much greater information is available in histories in sum-over-histories
quantum mechanics than is available on a spacelike surface.  We discuss
situations when this can become complete information about the system.

The spacetime approach to defining the complete information about a
closed quantum system reproduces the familiar (\ref{onetwo}) when
quantum theory can be formulated in terms of states on spacelike
surfaces. When quantum theory cannot be so formulated the spacetime
approach generalizes (\ref{onetwo}).  Section V describes this more
general notion of information in time-neutral
generalized quantum theories with initial and final conditions
\cite{Gri84,Har91a,GH93b}. The general case is discussed in Section VI
including generalized quantum theories of quantum spacetime
\cite{Harpp}. This is applied to spacetimes with closed, timelike curves
in Section VII.

Section VIII contains some brief remarks on the implications of the
spacetime approach to information for black hole evaporation.  When
quantum theory is not formulable in terms of states on a spacelike
surface complete information about a system is not necessarily to be
found on a given spacelike surface (even when ``spacelike'' can be defined).
Rather, one must search among all possible decoherent sets of spacetime
histories for those which give complete information.  In black hole
evaporation these may refer to alternatives on a spacelike surface after
the hole has evaporated as well as to alternatives near to or inside the
horizon.
The ``evolution of pure states into mixed states'' that is often
discussed \cite{Haw82}
 is thus not a violation of the principles of generalized
quantum mechanics.  It is only at conflict with the idea that the
evolution of the system can be completely described by states on
spacelike surfaces.

\section[]{Fixed Background Spacetimes}
\label{sec:II}

We begin by considering spacetime information in the approximation in
which spacetime geometry is fixed, foliable by spacelike surfaces,
 and quantum theory concerns
particles or fields moving in this given background.  This is an excellent
approximation on accessible scales for epochs
later than the Planck era.  As throughout this paper, we
consider a closed quantum system most generally the universe as a whole.
In this Section and in Sections III--IV
 we shall restrict attention to the usual formulation of quantum theory
in which probabilities for alternative, coarse-grained
histories are determined by a initial condition in the far past
represented by a density matrix $\rho$ together with an action or
Hamiltonian summarizing the dynamics of particles or fields
 in the fixed geometry.  We shall
consider generalizations of this standard framework in Sections V--VIII.  To
make our assumptions more precise, and to introduce the notation we
use, we now very briefly review the elements of the quantum
mechanics of closed systems.  We follow the treatment in
 \cite{GH90a,Har91a,GH93a} where more detailed
expositions as well as
references to the earlier literature may be found.

We assume that the fixed spacetime is foliable by spacelike surfaces
and pick a particular foliation, labeling
the surfaces by a time co\"ordinate, $t$.  A set of alternative
coarse-grained histories, of the closed system may be described by
giving sets of ``yes-no''
 alternatives at a sequence of times $t_1, \cdots, t_n$.
The alternatives at a particular time $t_k$ are represented by a
set of Heisenberg-picture projection operators
$\{P^k_{\alpha_k}(t_k)\}$.  In this notation, $\alpha_k$ is an index
specifying the particular alternative in the set and the superscript $k$
indicates that there may be different sets at different times. The
projections satisfy
\begin{equation}
P^k_{\alpha_k}(t_k) P^k_{\beta_k}(t_k) = \delta_{\alpha_k\beta_k}
P^k_{\alpha_k}(t_k)\ , \quad \sum_{\alpha_k} P^k_{\alpha_k} (t_k) = I
\label{twoone}
\end{equation}
which show that the alternatives are exclusive and exhaustive.
The operators $P^k_{\alpha_k}(t_k)$ obey the Heisenberg equation of
motion.
An individual history is
 a particular sequence of alternatives $(\alpha_1, \cdots,
\alpha_n)\equiv\alpha$ at the times $t_1, \cdots, t_n$.
 It is represented by the corresponding chain of
Heisenberg-picture projections:
\begin{equation}
C_\alpha = P^n_{\alpha_n}(t_n) \cdots P^1_{\alpha_1}(t_1)\ .
\label{twotwo}
\end{equation}
The set of all possible sequences gives a set of alternative histories
$\{C_\alpha\}$.
Evidently,
\begin{equation}
\sum\nolimits_\alpha C_\alpha = I\ .
\label{twothree}
\end{equation}

\noindent A set of histories $\{C_\alpha\}$
may be coarse-grained by partitioning it
into mutually
exclusive classes.  The class operators representing the individual
histories $\bar C_{\bar\alpha}$ in the coarser-grained set are the
sums of the $C_\alpha$ over the classes.
The general form of the class operators
representing an individual history in a set of exclusive alternative
ones is therefore
\begin{equation}
C_\alpha = \sum\limits_{(\alpha_1, \cdots, \alpha_n)\epsilon\alpha}
P^n_{\alpha_n}(t_n) \cdots P^1_{\alpha_1}(t_1)
\label{twofour}
\end{equation}
with (\ref{twothree}) continuing to hold.  Even more generally, the set
of histories may be branch dependent with the sets at later times
depending on the sets, times, and specific alternatives at earlier times
although we have not extended the notation for the $P$'s to indicate
this dependence explicitly.

The decoherence functional $D(\alpha^\prime, \alpha)$ assigns a complex
number to every pair of histories in a set of alternative ones
 that measures the coherence
between that pair.  It is defined by
\begin{equation}
D(\alpha^\prime, \alpha) = Tr \left(C_{\alpha^\prime} \rho
C^\dagger_\alpha\right)
\label{twofive}
\end{equation}
where $\rho$ is the Heisenberg picture density matrix representing the
initial condition of the closed system.  The set of histories is said to
decohere if the ``off-diagonal'' elements of $D(\alpha^\prime, \alpha)$
are sufficiently small. Quantum mechanics predicts probabilities only
for decoherent sets of histories whose probabilities obey the sum rules
of probability theory as a consequence of decoherence.
The probabilities $p(\alpha)$ of the individual
histories in a decoherent set are the diagonal elements of the
decoherence functional.

The probabilities of decohering sets of alternative histories provide
information about the system's initial condition.  The rest of this
section reviews the construction of this missing information by
M. Gell-Mann and the author \cite{GH90a,GHup}.
The missing information $S(\{C_\alpha\})$ in any particular set of histories
$\{C_\alpha\}$ is defined by employing a generalization of
the Jaynes ``maximum-entropy'' construction \cite{Ros83}. The entropy
functional of a density matrix
$\tilde\rho$ is defined to be
\begin{equation}
{\cal S}(\tilde\rho) = -Tr\left(\tilde\rho \log
\tilde\rho\right)\ .
\label{twosix}
\end{equation}
This is the standard information measure on density matrices; in Section
V we shall provide a justification of this formula from a more general
point of view. The
missing information $S(\{C_\alpha\})$ is the maximum of ${\cal
S}(\tilde\rho)$ over all density matrices $\tilde\rho$
 that contain the information available about the system through the
histories $\{C_\alpha\}$ and no more than that information. The information
available through $\{C_\alpha\}$ consists roughly of two parts, (i) the
decoherence of the set and (ii) the probabilities of the individual
histories in the set.  A $\tilde\rho$ that reproduces this information
 should reproduce the
decoherence
functional for the set of histories $\{C_\alpha\}$:
\begin{equation}
Tr\left(C_{\alpha^\prime} \tilde\rho C^\dagger_\alpha\right) = Tr
\left(C_{\alpha^\prime}\rho C^\dagger_\alpha\right)\ .
\label{twoseven}
\end{equation}
Thus,
\begin{equation}
S(\{C_\alpha\}) = \mathop{\rm max}_{\tilde\rho}
\Bigl[{\cal S}(\tilde\rho)\Bigr]_{Tr\left(C_\alpha\tilde\rho
C^\dagger_\alpha\right) = D\left(\alpha^\prime, \alpha\right)}\ .
\label{twoeight}
\end{equation}
The definition (\ref{twoeight}) incorporates any standard of approximate
decoherence that may be enforced.  If the off-diagonal elements
of $D(\alpha^\prime,\alpha)$
are required to be zero to some accuracy, the density matrix
$\tilde\rho_{\rm max}$ that determines ${\cal S}(\{C_\alpha\})$ will
reproduce decoherence and probabilities\footnote{It would be
possible to define a missing information in the probabilities alone
simply by reproducing only the diagonal elements of $D$ in
(\ref{twoeight}).
with that accuracy for histories $\{C_\alpha\}$.
The missing information so defined would, of course,
be greater than that including decoherence.}
with that accuracy for histories $\{C_\alpha\}$.
The definition
(\ref{twoeight}) is consistent with the notion of physically equivalent
sets of histories described in \cite{GH94}. Sets of histories whose class
operators $\{C_\alpha\}$ differ by a reassignment of the times in
(\ref{twofour}) or by a constant unitary transformation that preserves
the initial $\rho$ are physically equivalent.  Since physically
equivalent sets have the same decoherence functional they will also have
the same missing information through (\ref{twoeight}).

The density matrix $\tilde\rho_{\rm max}$ that maximizes
${\cal S}(\tilde\rho)$
subject to the constraints (\ref{twoseven}) may be found by the method
of Lagrange multipliers. One first extremizes ${\cal S}(\tilde\rho)$
with respect to all operators $\tilde\rho$ (not just density matrices)
enforcing (\ref{twoseven}). The result is
\begin{equation}
\tilde\rho_{\rm max} =  \exp \left(-\sum\nolimits_{\alpha\alpha^\prime}
\lambda^{\alpha\alpha^\prime} C^\dagger_\alpha
C_{\alpha^\prime}\right)\ .
\label{twonine}
\end{equation}
The Lagrange multipliers $\lambda^{\alpha\alpha^\prime}$ are determined
by (\ref{twoseven}) and there is a solution with
$\lambda^{\alpha\alpha^\prime} = (\lambda^{\alpha^\prime\alpha})^*$. Thus
$\tilde\rho_{\rm max}$ is Hermitian.  It is also normalized because
summing both sides of (\ref{twoseven}) over $\alpha$ and $\alpha^\prime$
gives $Tr(\tilde\rho_{\rm max}) = Tr(\rho)=1$. The operator
$\tilde\rho_{\rm max}$ is therefore a density
matrix.  The missing information
$S(\{C_\alpha\})$ is easily expressed directly in terms of the
multipliers and the decoherence functional
\begin{equation}
S(\{C_\alpha\}) = \sum\nolimits_{\alpha\alpha^\prime}
\lambda^{\alpha\alpha^\prime} D(\alpha^\prime, \alpha) \approx
\sum\nolimits_\alpha \lambda^{\alpha\alpha} p(\alpha)\ ,
\label{twoten}
\end{equation}
the last equality holding when the set $\{C_\alpha\}$ decoheres.

The conditions (\ref{twoseven}) are in general difficult to solve for
the multipliers $\lambda^{\alpha\alpha^\prime}$, but there is one
useful case where the solution may be obtained explicitly.  That is when the
histories consist of alternatives at a single moment of time so that the
$\{C_\alpha\}$ are a set of orthogonal projections $\{P_\alpha\}$.
Decoherence is then automatic from the cyclic property of the trace in
(\ref{twofive}) and the orthogonality of the $\{P_\alpha\}$ [{\it
cf.} (\ref{twoone})].
The conditions (\ref{twoseven}) reduce
to
\begin{equation}
p(\alpha) = Tr \left[P_\alpha \exp \left(-\Sigma_\beta \lambda^\beta
P_\beta\right)\right]
\label{twoeleven}
\end{equation}
whose solution is
\begin{equation}
\lambda^\alpha = - \log \left[p(\alpha)/Tr\left(P_\alpha\right)\right]\
{}.
\label{twotwelve}
\end{equation}
The result for $\tilde\rho_{\rm max}$ in (\ref{twonine}) is
\begin{equation}
\tilde \rho_{\rm max} = \sum\nolimits_\alpha\left[p(\alpha)
P_\alpha/Tr(P_\alpha)\right]\ ,
\label{twothirteen}
\end{equation}
and then
\begin{equation}
S(\{P_\alpha\}) = - \sum\nolimits_\alpha p(\alpha)\log p(\alpha)
+ \sum\nolimits_\alpha p (\alpha) \log Tr(P_\alpha)\ .
\label{twofourteen}
\end{equation}

In simple cases the expression (\ref{twofourteen}) gives the familiar
entropy of statistical mechanics \cite{Ros83,GHup}. Suppose $\rho$ is
an eigenstate of the total energy --- the ``microcanonical ensemble''
and let $\{P_\alpha\}$ be projections down on ranges of energy of width
$\Delta E$.  Then clearly $p(\alpha)=0$ for all ranges except that
containing the energy of $\rho$ and so
\begin{equation}
S\left(\{P_\alpha\}\right) = \log\ Tr\left(P_\alpha\right) = \log
\ N \label{twofifteen}
\end{equation}
where $N$ is the number of states with energy $E$ in the range $\Delta
E$. This is the usual Boltzmann entropy.

The missing information in any coarse grainings of the set $\{C_\alpha\}$ is
greater than the missing
information in $\{C_\alpha\}$ itself.  To see this note
that the class operators for a coarser-grained set $\{\bar C_{\bar\alpha}\}$
are
{\it sums} of the class operators for $\{C_\alpha\}$:
\begin{equation}
\bar C_{\bar\alpha} = \sum\limits_{\alpha\epsilon\bar\alpha}
C_\alpha\ .
\label{twosixteen}
\end{equation}
Correspondingly, the conditions (\ref{twoseven}) for $\{\bar C_{\bar\alpha}\}$
are sums of the conditions for $\{C_\alpha\}$.  They are therefore
{\it weaker} conditions, and the maximum in (\ref{twoeight}) can only be
greater for $\{\bar C_{\bar\alpha}\}$ than it is for $\{C_\alpha\}$.  Thus,
\begin{equation}
S\left(\{\bar C_{\bar\alpha}\}\right) \geq S \Bigl(\{C_\alpha\}\Bigr)
\ .\label{twoseventeen}
\end{equation}

Requiring that the density matrix $\tilde\rho$ reproduce the full
decoherence functional $\rho$ rather than, say, just the diagonal elements
which are the probabilities means that $\tilde\rho$ functions as an
initial condition not only for $\{C_\alpha\}$ but for all coarser
grainings of it.  In particular the probabilities for coarser-grained
sets of histories are given by
\begin{equation}
p(\bar\alpha) = Tr\left(\bar C_{\bar\alpha} \tilde\rho \bar
C^\dagger_{\bar\alpha}\right)
\label{twoeighteen}
\end{equation}
because the set $\{\bar C_{\bar\alpha}\}$ decoheres with respect to
$\tilde\rho$.

We have defined the missing information in a particular set of
alternative histories $\{C_\alpha\}$.  Given a class of sets of
histories, one member may yield more information about the system than
another.  We therefore define the missing information in a class ${\cal
C}$ of sets of alternative decohering histories as
\begin{equation}
S({\cal C}) = \mathop{\rm min}_{\{C_\alpha\}\epsilon{\cal C}}
S\Bigl(\{C_\alpha\}\Bigr)\ .
\label{twonineteen}
\end{equation}

Computing $S({\cal C})$ for different classes
enables one to understand different ways information about a quantum
system can be obtained.
For example we could study whether the same information
is available in sum-over-histories quantum mechanics as using
a general operator formulation,
and whether the same information is available in
homogeneous histories\footnote{Following the terminology of Isham
\cite{Ish94}.}, which are chains of $P$'s as in (\ref{twotwo}), {\it vs.}
the more general inhomogeneous histories which are sums of chains as in
(\ref{twofour}).

The minimum of $S(\{C_\alpha\})$ over the class of
{\it  all} decohering histories is the
least missing information about the system --- the complete information.
We write this as $S_{\rm compl}$  without an
argument
\begin{equation}
S_{\rm compl} =\mathop{\min}_{{\rm decohering}\atop \{C_\alpha\}}
S\Bigl(\{C_\alpha\}\Bigr)\ .
\label{twotwenty}
\end{equation}

In the usual quantum mechanics under discussion,
no more information is available about the system in any set of
histories than is contained in the initial density matrix by the measure
(\ref{twosix}). To see this
first note the general inequality\footnote{For a convenient proof see
\cite{Rue69}.} for any pair of density matrices $\rho_1$ and $\rho_2$.
\begin{equation}
-Tr\left(\rho_1 \log \rho_2\right) \geq - Tr\left(\rho_1 \log
\rho_1\right)\ .
\label{twotwentyone}
\end{equation}
Then note that from the explicit form (\ref{twonine}) it follows
that\footnote{This is one of the requirements that $\tilde\rho_{\rm
max}$ be a coarse-graining of $\rho$, see \cite{GHup}.}
\begin{equation}
S(\{C_\alpha\}) \equiv -Tr\left(\tilde\rho_{\rm max}\log \tilde\rho_{\rm
max}\right) = -Tr\left(\rho\log \tilde\rho_{\rm max}\right)\ .
\label{twotwentytwo}
\end{equation}
Finally, using (\ref{twotwentyone}), we have
\begin{equation}
S(\{C_\alpha\}) \geq {\cal S}(\rho)\ .
\label{twotwentythree}
\end{equation}

The bound (\ref{twotwentythree}) is realized for a set of histories in
which $C_\alpha=P_\alpha$ where the $P_\alpha$ are projections onto a
complete set of states that diagonalize $\rho$. That is because
decoherence is automatic for sets of histories that are
projections, as a consequence of the cyclic property of the trace in
(\ref{twofive}). Thus,
\begin{equation}
S_{\rm compl}={\cal S}(\rho)\equiv -Tr(\rho \log \rho) \  .
\label{twotwentythreea}
\end{equation}
Since in the Heisenberg picture every operator
corresponds to some quantity at any time the $\{P_\alpha\}$ may be regarded
as residing on any spacelike surface. Thus, on a spacelike surface it is
always possible to recover complete information, $S_{\rm compl}$, about the
initial
condition through a
suitable choice of decoherent histories consisting of a single set of
projections.

\section[]{Information in Spacetime Regions}
\label{sec:III}

An interesting and useful example of the missing information in a class
of histories is the missing information in the class of histories that
refer only to a particular spacetime region. To
illustrate the idea we consider a quantum field theory with a single
scalar field $\phi(x)$.  The generalization to a realistic panoply of
gauge, spinor, and tensor fields should be straightforward.  Given a
spacetime region $R$ at time $t$, we can define operators ${\cal
O}(R,t)$ that refer only to the region $R$ as
  functions of the fields and
their conjugate momenta {\it inside} $R$ at that time, {\it viz}.
\begin{equation}
{\cal O}(R, t) = {\cal O} \bigr(\phi({\bf x}, t), \pi ({\bf x}, t)\bigr)
\,, \quad {\bf x} \epsilon R\ {\rm at}\ t\ .
\label{threeone}
\end{equation}
Alternative range of values of such variables are represented by sets
of projection operators.  We denote these by $\{P_\alpha^{{\cal O}(R,t)}
(t)\}$ where the discrete index $\alpha$ runs over an exhaustive set of
mutually exclusive ranges of ${\bf R}$. (The two occurences of $t$ are
redundant; we include them for clarity.)
Sequences of such sets of alternatives ${\cal O}_1(R,t_1), \cdots,
{\cal O}_n(R,t_n)$ at a series of times $t_1, \cdots
, t_n$ that are contained within the span of $R$ define examples of sets of
alternative histories that refer only to $R$.  Individual histories
correspond to particular sequences of alternatives $\alpha \equiv
(\alpha_1, \cdots, \alpha_n)$ at the definite moments of time and are
represented by the corresponding chain of projection operators
\begin{equation}
C_\alpha(R) = P^{{\cal O}_n(R,t_n)}_{\alpha_n}(t_n)
\cdots P^{{\cal O}_1(R,t_1)}_{\alpha_1} (t_1)\ .
\label{threetwo}
\end{equation}

More generally, we may consider partitions of such histories into
mutually exclusive classes $\{c_\alpha\}$ and consider the limit where
there is one alternative at each and every time. The resulting class
operators have the form
\begin{equation}
C_\alpha(R) = {\mathop{\lim}_{n\to\infty}} \sum\limits_{(\alpha_1\cdots
\alpha_n)\epsilon\alpha} P^{{\cal O}_n (R,t_n)}_{\alpha_n}(t_n) \cdots P^{{\cal
O}_1(R,t_1)}_{\alpha_1} (t_1) \ .
\label{threethree}
\end{equation}
This is the most general notion of a set of
 alternative histories that refers to
a spacetime region $R$, and operators of this form define the class of
histories ${\cal C}_R$ that refer only to the spacetime region $R$.

A simple example may be helpful.  Consider the average of the field
$\phi(x)$ over the region $R$:
\begin{equation}
\phi(R) = {1\over V(R)} \int\nolimits_R d^4 x \phi(x)
\label{threefour}
\end{equation}
where $V(R)$ is the four-dimensional volume of $R$.  Of course, this
average can be written
\begin{equation}
\phi(R) = {1\over V(R)} \int\nolimits_R dt\, \phi (R, t)
\label{threefive}
\end{equation}
where by $\phi(R, t)$ we mean the {\it spatial} average over the
intersection  of $R$ with the constant time surface labeled by $t$.
Take $\phi(R,
t)$ for the ${\cal O}(R, t)$ at a discrete series of times $t_1, \cdots,
t_n$ equally spaced by a small interval $\epsilon$. Take a set of
small intervals of ${\bf R}$ of equal
size $d$ for the ranges labeled by $\alpha$.  Then the history in
which the space{\it time} average
$\phi(R)$ lies in the range $\Delta$ is represented by a sum of the form
(\ref{threethree}).  The sum is over all ranges $\alpha_1, \cdots,
\alpha_n$ of $\phi(R, t_1), \cdots, \phi(R, t_n)$ such that the their
central values, $\phi_{\alpha_1}, \cdots, \phi_{\alpha_n}$ satisfy
\begin{equation}
{1\over V(R)} \sum\nolimits_k \epsilon\phi_{\alpha_k}\ \in
\ \Delta\ .
\label{threesix}
\end{equation}
As $\epsilon\to 0$, $n\to\infty$, and $d\to 0$ the formula
(\ref{threethree}) gives the class
operator\footnote{Note that this representation of $\phi(R)$ in the
range $\Delta$ is not the projection operator of the average of
Heisenberg fields onto the range $\Delta$.  The class operator defined
by (\ref{threesix}) is not
generally a projection operator.  Ranges of values of time averages of
Heisenberg operators certainly describe other alternatives, but to be
incorporated into a quantum framework that deals with histories they
would have to be assigned a time ({\it cf.} \cite{Har91a}, Section IV.1).
That is certainly possible since every Heisenberg picture projection may
be interpreted as the projection into the value of some quantity at any
time.  However, when the region $R$ extends over time there is
 no natural value for this time.
 The present construction does not require such a
specification and corresponds to the sum-over-histories definition of
alternatives as partitions of field histories (see {\it e.g.} \cite{Harpp}).
The different operator representations of the same classical
quantity reflects the usual factor ordering ambiguity in quantum
mechanics, arising in this case because field values at different times
generally do not commute.}
corresponding to the history in which $\phi(R)$ lies in the
range $\Delta$.

We can now define the missing information, $S(R)$ associated with a
spacetime region $R$ as the minimum of the missing information in
the class ${\cal C}_R$ of decoherent sets of histories $\{C(R)_\alpha\}$
that refer only to $R$. In symbols,
\begin{equation}
S(R) = \mathop{\min}_{{\rm decohering}\atop\{C_\alpha(R)\}}
S\left(\{C_\alpha(R)\}\right)\ .
\label{threeseven}
\end{equation}
Thus $S(R)$ is a measure of how far we are from having complete
information about a system if we only have access to alternatives inside
a spacetime region $R$.  Since $S(R)$ is defined for space{\it time}
regions it is a fully four-dimensional notion of information.

We can illustrate this idea with three examples shown in Figure 1. The
region in Fig.~1a contains a spacelike surface $\sigma$. Complete
information is therefore available through the set of histories
represented by projections onto the basis which diagonalizes
$\rho(\sigma)$.  The missing information is
\begin{equation}
S(R_a) = - Tr(\rho\log\rho)\ .
\label{threeeight}
\end{equation}
Complete information cannot be
available for regions like those of Fig 1b
 because the initial state may contain wave packets that never
cross $R$, at least if it has a finite extent in the time direction.
Thus
\begin{equation}
S(R_b) \geq  - Tr(\rho\log\rho)\ .
\label{threenine}
\end{equation}
The question would become more interesting if space were closed so that
any  wave packet heading away from $R_b$ would inevitably return.

In Figure 1c the region $R$ is the domain of dependence of a region $L$
of a spacelike surface $\sigma$.  The field equations (or equivalently
the Heisenberg equations of motion) permit every $\phi(x),\ x\in R$
to be expressed in terms of $\phi(x),\ x\in L$ on the spacelike
surface.  Thus every $\{C_\alpha (R)\}$ can be so reexpressed.  Further,
every $\{C_\alpha(L)\}$  certainly  refers to the spacetime region $R$.
Thus the missing information in $R$ is the same as the missing
information in the region $L$ of $\sigma$, $S(R_c)=S(L)$.
The missing information $S(L)$ may be calculated if we assume a
regularization of the field degrees of freedom so that it is meaningful
to speak of a Hilbert space ${\cal H}_L \otimes {\cal H}_{\bar L}$ that
is a tensor product of a factor for the degrees of freedom inside $L$
an another for those outside. $S(L)$ is then the minimum of
(\ref{twofourteen}) over sets of projections of the form $P_\alpha
\otimes I$ that refer only to the region inside $L$. A straightforward
calculation shows:
\begin{equation}
S(R_c)=S(L) = - tr\bigl[Sp(\rho) \log (Sp(\rho)/Sp(I))\bigr]
\label{threeten}\\
\end{equation}
where $Sp$ denotes the trace over ${\cal H}_{\bar L}$ and $tr$ is the
trace over ${\cal H}_L$. Of course, from the general result
(\ref{twotwentythree})
\begin{equation}
S(R_c)=S(L) \geq -Tr\bigl(\rho\log\rho\bigr)\ .
\label{threetena}
\end{equation}

The notion of missing information in a spacetime region $R$ is a fully
four-dimensional way of discussing the localization of information in a
fixed background spacetime.  In Section VIII we shall apply it to a
discussion of information in black hole spacetimes.

\section[]{Information in Sum-Over-Histories Quantum Mechanics}
\label{sec:IV}

Feynman's sum-over-histories formulation of quantum mechanics is an
alternative, spacetime, formulation of quantum theory.  It agrees with
the usual quantum theory formulated in terms of states, operators,
etc.~for alternatives that can be described in terms of the
co\"ordinates of configuration space.  However, it differs from usual
quantum theory in that it is restricted to such configuration space
alternatives. Alternative values of momentum, for example, are not
defined at an instant of time, but, only approximately, in terms of
configuration space alternatives at several moments of time \cite{FH65}.
  It is,
therefore, an interesting question whether complete information, in the
sense we have defined it, is
available in sum-over-histories quantum mechanics
 from its more limited class of alternatives.  We examine that
question in this section.

The missing information in sum-over-histories quantum mechanics is
defined, as in (\ref{twonineteen}), as the minimum of the missing
information in histories of sum-over-histories form, {\it viz.},
\begin{equation}
S({\rm soh}) = \mathop{\min}_{{\rm
decohering}\atop \{C_\alpha\}\in {\cal C}_{\rm soh}}
S\Bigl(\{C_\alpha\}\Bigr)\ .
\label{fourone}
\end{equation}
To understand more precisely what this formula means, and to take some
first steps in its evaluation, let us consider the model of a single,
non-relativistic particle  moving in $\nu-$dimensions.

The fine-grained histories of a non-relativistic particle are paths
$q^i(t)$ which are single-valued functions of the time, $t$, say, on an
interval $[0, T]$.  The general, spacetime notion of a set of
alternatives for which quantum theory might predict probabilities is a
partition of this set of fine-grained paths into mutually exclusive
classes $\{c_\alpha\}$, $\alpha=1, 2, \cdots$\cite{Har91b,YT92}. The
totality of all such partitions defines the class of histories of
sum-over-histories form ${\cal C}_{\rm soh}$. The class operators
representing such alternatives have
matrix elements given by
\begin{equation}
\bigl\langle q^{\prime\prime} | C_\alpha |
q^\prime\bigr\rangle = \int\nolimits_{[q^\prime c_\alpha q^{\prime\prime}]}
\delta q e^{iS[q(\tau)]}\ .
\label{fourtwo}
\end{equation}
Here, $S[q(\tau)]$ is the action functional, units are used where
$\hbar=1$, and  the sum is over paths which
start at $q^\prime$ at $t=0$, end at $q^{\prime\prime}$ at $t=T$, and
lie in the class $c_\alpha$. Co\"ordinate indices have been expressed
for compactness. As shown by Caves and others
\cite{Cav86,Sta86,YT92,Har91b}
the operators $\{C_\alpha\}$ may be expressed as a limit of
forms like (\ref{twofour}) where the times become dense in the interval
$[0, T]$ and the projections are onto ranges of position.
The decoherence functional is given by (\ref{twofive}) and
the missing information in a set $\{C_\alpha\}$ by
(\ref{twoeight}).  The missing information in all histories of
sum-over-histories form is then given by (\ref{fourone}) where the
minimum is over partitions of the paths $\{c_\alpha\}$ that decohere.

As the simplest example, consider a partition of paths by which of a set of
intervals $\{\Delta^1_\alpha\}$ the particle
passes through at time $t_1$.  The path
integral (\ref{fourtwo}) can be rewritten
\begin{equation}
\bigl\langle q^{\prime\prime} | C_\alpha |
q^\prime\bigr\rangle = \int\nolimits_{\Delta^1_\alpha} dq
\left(\int\nolimits_{[q, q^{\prime\prime}]} \delta q\,
e^{iS[q(\tau)]}\right)
\left(\int\nolimits_{[q^\prime, q]} \delta q\,
e^{iS[q(\tau)]}\right)
\label{fourthree}
\end{equation}
where the individual path integrals in (\ref{fourthree}) are over
unrestricted paths from $(q',0)$ to $(q,t_1)$ and $(q,t_1)$ to
$(q^{\prime\prime}, T)$ respectively. The unrestricted path integral
from $(q_1, t_1)$ to $(q_2, t_2)$ defines a propagator according to
\begin{equation}
\int\nolimits_{[q_1, q_2]} \delta q\, e^{iS[q(\tau)]} = \langle q_2
t_2 | q_1 t_1\rangle = \langle q_2 |
e^{-iH(t_2-t_1)}|q_1\rangle\ .
\label{fourfour}
\end{equation}
Combining (\ref{fourthree}) and (\ref{fourfour}) one has
\begin{equation}
C_\alpha = e^{-iHT} P^1_\alpha(t_1)\ ,
\label{fourfive}
\end{equation}
where $P^1_{\alpha_1}(t_1)$ is the Heisenberg-picture projector onto the
region $\Delta^1_\alpha$ at time $t_1$. [We are thus, in this section,
using a convenient normalization of the class operators such that
$\Sigma_\alpha C_\alpha = \exp (-iHT)$ rather than (\ref{twothree}). The
value of the decoherence functional is unaffected by this choice.]

$S(\{C_\alpha\})$ for histories consisting of a projection at a single
moment of time was calculated in eq (\ref{twofourteen}).
$S(\{C_\alpha\})$ for the histories (\ref{fourfive}) is the same since
the factor $\exp(-iHT)$ does not affect the value of the decoherence
functional in (\ref{twofive}) by the cyclic property of the trace. The
probabilities $p_\alpha= Tr[P^1_\alpha(t_1)\rho]$ will be finite for
reasonable $\rho$, but the traces of the projection operators in
(\ref{twofourteen}) will diverge
\begin{eqnarray}
Tr[P^1_\alpha(t_1)] &=& \int dr_1 \langle
r_1t_1|P^1_\alpha(t_1) | r_1t_1\rangle\nonumber\\
&=& \int\nolimits_{\Delta^1_\alpha} dq_1 \int dr_1\, \left\langle r_1t_1|
q_1t_1\right\rangle\left\langle q_1t_1 | r_1t_1\right\rangle\nonumber\\
&=& \delta^{(\nu)} (0) V(\Delta^1_\alpha) = \infty
\label{foursix}
\end{eqnarray}
where $V(\Delta)$ is the configuration-space
volume of the region $\Delta$.  Thus, there is an
infinite amount of missing information in histories that are partitions
of  the
paths at a single moment of time.

The situation is already much improved with two times.  Suppose the
paths are partitioned by their behavior with respect to sets of regions
$\{\Delta^1_{\alpha_1}\}$ at time $t_1$ and $\{\Delta^2_{\alpha_2}\}$ at
time $t_2$.  By the straightforward extension of the argument given
above, the class operators are
\begin{equation}
C_\alpha = e^{-iHT} P^2_{\alpha_2} (t_2) P^1_{\alpha_1}
(t_1)
\label{fourseven}
\end{equation}
where the projections are onto the regions $\Delta^2_{\alpha_2}$ at
$t_2$ and $\Delta^1_{\alpha_1}$ at $t_1$ respectively.  The missing
information in such a set is no longer given by the simple formula
(\ref{twofourteen}) but now must by calculated from (\ref{twoten}). The
$\lambda^{\alpha\alpha^\prime}$ are determined by the condition.
\begin{equation}
D\left(\alpha^\prime, \alpha\right) =
Tr\left[C^\dagger_\alpha C_{\alpha^\prime}
\exp\left(-\sum\nolimits_{\beta\beta^\prime}\lambda^{\beta\beta^\prime}
C^\dagger_\beta C_{\beta^\prime}\right)\right]
\label{foureight}
\end{equation}
with $C$'s of the form (\ref{fourseven}).  Expanding the exponential in
a power series one finds a series of terms with coefficients of the form
\begin{equation}
Tr \left[P^1_{\mu_1}(t_1) P^2_{\mu_2} (t_2) P^1_{\mu_3}(t_1)
P^2_{\mu_4} (t_2) \cdots P^2_{\mu_m}(t_2)\right]\ ,
\label{fournine}
\end{equation}
that is,
strings of alternating $P^1(t_1)$'s and $P^2(t_2)$'s. These
traces are all finite. For example, the simplest  one is
\begin{equation}
Tr\left[P^1_{\mu_1}(t_1) P^2_{\mu_2}(t_2)\right] =
\int\nolimits_{\Delta^1_{\mu_1}} dq_1 \int\nolimits_{\Delta^2_{\mu_2}} dq_2
\left\langle q_1t_1|q_2t_2\right\rangle
\left\langle q_2t_2|q_1t_1\right\rangle\ .
\label{fourten}
\end{equation}
The propagators are not divergent if $t_1\not= t_2$ and the
$q$-integrals are over finite ranges.  For example, if the propagators were
those of a free particle of mass $m$ in $\nu-$dimensions
\begin{equation}
Tr\left[P^1_{\mu_1}(t_1) P^2_{\mu_2}(t_2)\right] =
\frac{V(\Delta^1_{\mu_1})
V(\Delta^2_{\mu_2})}{\left[(2\pi/m)i|t_1-t_2|\right]^\nu}\ .
\label{foureleven}
\end{equation}
Of course, as $t_2\to t_1$, the propagators diverge and we recover the
infinity of (\ref{foursix}). Mere finiteness of the coefficients in
the expansion of the equation  [(\ref{foureight})] that determines the
multipliers $\lambda^{\alpha\alpha^\prime}$ does not imply that the
solutions will be finite but is at least consistent with it.  By
contrast a similar expansion of (\ref{twoeleven}) in the case of
alternatives at a single time yields divergent coefficients and a
divergent solution (\ref{twotwelve}).

It is therefore plausible that, while the missing information in
partitions of paths at a single time is infinite, the missing
information in partitions that involve several times is finite.  There
is thus an infinite improvement in passing from a single time to many.
That is perhaps not so very surprising.  The alternatives usually
defined by momentum operators, for example,
 are not available at  a single  moment of
time in sum-over-histories quantum mechanics.  However, they are
available approximately through models of time of flight determinations
of momentum involving two or more times \cite{FH65}.

We are thus led to the interesting question of whether the complete
information in sum-over-histories quantum mechanics is greater than that
in the usual operator formulation or coincides with it.  Even if they
differ, the above arguments suggest that they differ by only a finite
amount for finite dimensional configuration spaces.\footnote{A.~Connes
(private communication)
has shown that the two notions of complete information do not coincide for some
simple models with finite dimensional Hilbert
spaces, where restrictions to projections on a particular basis is the
analog of the restriction to configuration space histories in a
sum-over-histories formulation.} The value of this finite difference would
itself be interesting.

It is already known that the decoherence functional for general
alternatives can be recovered from the sum-over-histories decoherence
functional through suitable transformations \cite{Linup}.
 Were the complete information
available in sum-over-histories quantum mechanics the same as in the
operator versions of the theory, that would be another argument for the
sufficiency of a sum-over-histories formulation for prediction in
physics.

\section[]{Information in Time-Neutral Generalized \\  Quantum Mechanics}
\label{sec:V}

\subsection[]{Quantum Mechanics with Initial and Final Conditions}
\label{subsec:va}

We now turn to information in generalized quantum theories
\cite{Har91a,Harpp,Ish94}.  A general discussion will be given in the
next section, but as an introduction we consider one of
the simplest examples in this section --- quantum mechanics
with both initial and final
conditions \cite{ABL64a,Gri84,Har91a,GH93b}. This is a quantum theory
whose notions fine- and coarse-grained histories coincide with those of
  the  usual formulation in Section II.
Individual members of a set of alternative histories continue to be
represented by
class operators $C_\alpha$ of the general form (\ref{twofour})
in a Hilbert space ${\cal H}$. Only the
decoherence functional differs from the usual (\ref{twofive}) by
incorporating both an initial condition represented by a positive
Hermitian matrix $\rho^i$ and a final condition represented by a
positive Hermitian matrix $\rho^f$.  This decoherence functional is
\begin{mathletters}
\begin{equation}
D(\alpha^\prime, \alpha) = {\cal N}\ Tr\left(\rho^fC_{\alpha^\prime}
\rho^i C^\dagger_\alpha\right)\ ,
\label{fiveonea}
\end{equation}
where ${\cal N}$ is determined so that $\Sigma_{\alpha^\prime\alpha}
D(\alpha^\prime, \alpha) =1$, specifically,
\begin{equation}
{\cal N}^{-1} = Tr\left(\rho^f\rho^i\right)\ .
\label{fiveoneb}
\end{equation}
\label{fiveone}
\end{mathletters}

\noindent A set of histories is said to (medium) decohere when the off-diagonal
elements of (\ref{fiveonea}) are sufficiently small; the
approximate probabilities of the histories are then the diagonal
elements.  These probabilities are consistent with the rules of
probability theory as a consequence of decoherence.

Quantum theory based on the decoherence functional (\ref{fiveone}) is a
generalization of usual quantum theory described in Section II. The decoherence
functional
of the usual formulation (\ref{twofive}) is the special case of
(\ref{fiveone}) with $\rho^f=I$ and $\rho^i=\rho$.  In contrast
to the usual formulation, which incorporates an arrow of time, the
generalized quantum theory based on
 (\ref{fiveone}) is time
neutral.  The decoherence functional (\ref{twofive}) distinguishes the
ends of the histories.  At one end (conventionally called the past)
there is a density matrix.  At the other end (conventionally called the
future) there is the trace. The
generalized form (\ref{fiveone}) treats the ends symmetrically
and $\rho^f$ and $\rho^i$ can be interchanged using the cyclic properties
of the trace.  In a quantum cosmology based on the time-neutral
 (\ref{fiveone})
and time-symmetric dynamical laws, all observed arrows of time
arise from differences between $\rho^i$ and $\rho^f$ \cite{GH93b}.

Usual quantum
mechanics can be formulated in terms of an evolving state on a spacelike
surface that summarizes the past for the purposes of future prediction
-- an essentially time-asymmetric notion.  Clearly, time-neutral
generalized quantum theory
cannot be so formulated, but is fully predictive as we have
described.  For a fuller discussion of this and other features of time-neutral
generalized quantum theory see
\cite{Gri84,GH93b}.  We shall now discuss the appropriate
generalizations of the notions of information that were
described for the usual
theory in Section II.

\subsection[]{Information in the Initial and Final Conditions}
\label{subsec:vb}

In Section II we defined the missing information in a set of histories
making use of the information measure ${\cal S} (\tilde\rho) =
-Tr(\tilde\rho \log \tilde\rho)$ on density matrices. We
posited this measure; we did not derive it.  To define the analogous
notions of information in time-neutral generalized quantum mechanics we
need a measure of information in {\it pairs} of positive, Hermitian
operators $\tilde\rho^i$ and $\tilde\rho^f$.  We now derive that
measure.

A general approach to the definition is to define the missing
information ${\cal S} (\tilde\rho^f, \tilde\rho^i)$ in
$\tilde\rho^i$ and $\tilde\rho^f$ as $-\Sigma_\alpha[p(\alpha)\log p
(\alpha)]$ for some set of probabilities $\{p(\alpha)\}$ determined by
$\tilde\rho^i$ and $\tilde\rho^f$. These probabilities are naturally the
probabilities of some set of decoherent histories as determined by the
decoherence functional (\ref{fiveone}).  However, we cannot
use histories that are too coarse grained or the measure will be
trivial.  For example, if we use the maximally coarse grained set
consisting of the single history $P=I$, then it has probability 1 and $-p
\, \log p=0$. Put more informally, the form $-\Sigma_\alpha [p(\alpha)\, \log
p(\alpha)]$ contains no penalty for asking stupid questions.  Rather, in
order to
define the missing information in $\tilde\rho^i$ and
$\tilde\rho^f$, we should consider the probabilities of only some
standard class of very fine-grained decoherent sets.  The natural
candidate for this standard class in the case of time-neutral quantum
mechanics is the class of completely fine-grained decoherent sets, since
such sets exist, as we shall now show.  (We shall return to a discussion
of this standard class in the general discussion in Section VI.)  We thus
define
\begin{equation}
{\cal S} \bigl(\tilde\rho^f, \tilde\rho^i\bigr) \equiv
\mathop{\min}_{{\rm fine-grained}\atop{\rm decoherent}
\ \{C_\alpha\}}
\left[-\sum\nolimits_\alpha p(\alpha)\log p(\alpha)\right]
\label{fivetwo}
\end{equation}
where the minimum is over the fine-grained decoherent
 sets $\{C_\alpha\}$ for which
\begin{equation}
D(\alpha^\prime, \alpha) = {\cal N}\ Tr\left(\tilde\rho^f
C_{\alpha^\prime} \tilde\rho^i C^\dagger_\alpha\right) =
\delta_{\alpha^\prime\alpha} p (\alpha)\ .
\label{fivethree}
\end{equation}
We now compute the measure ${\cal S}(\tilde \rho^f, \tilde\rho^i)$ so defined
 as an explicit
functional of $\tilde\rho^f$ and $\tilde\rho^i$.

Fine-grained histories consist of sequences of sets of one-dimensional
projections onto a basis for ${\cal H}$ (a complete set of states)  at
each and every time.  To keep the notation manageable let us consider
for a moment a finite sequence of times $t_1, \cdots, t_n$. The $\{C_\alpha\}$
representing the finest-grained histories at these times may be written
\begin{equation}
C_\alpha = P^n_{\alpha_n} \cdots P^1_{\alpha_1}, \qquad
 P^k_{\alpha_k} = | k,
\alpha_k \rangle \langle k, \alpha_k |
\label{fivefour}
\end{equation}
where $\{| k, \alpha_k\rangle\}$ are a set of basis vectors at time $t_k$
as $\alpha_k$ ranges over a set of discrete indices.  To compress the
notation we shall write $\{|\alpha_k\rangle\}$ for these basis vectors at
each time, remembering that there may be different sets at different
times.

The condition that a set of the form (\ref{fivefour}) decoheres is, from
(\ref{fivethree}),
\begin{equation}
\langle\alpha_n|\tilde\rho^f|\alpha^\prime_n\rangle
\langle\alpha^\prime_n|\alpha^\prime_{n-1}\rangle \cdots
\langle\alpha^\prime_2 |\alpha^\prime_1\rangle
\langle\alpha^\prime_1 | \tilde\rho^i |
\alpha_1\rangle\langle\alpha_1|\alpha_2\rangle \cdots
\langle\alpha_{n-1} | \alpha_n\rangle = 0
\label{fivefive}
\end{equation}
whenever any $\alpha^\prime_k \not= \alpha_k$. The probabilities of the
individual histories in this decoherent set are
\begin{equation}
p\left(\alpha_n, \cdots, \alpha_1\right) = \bigl\langle\alpha_n |
\tilde\rho^f|\alpha_n\bigr\rangle\, \left | \left\langle\alpha_n | \alpha_{n-1}
\right\rangle\right |^2 \cdots \left |\left\langle\alpha_2 |
\alpha_1\right\rangle\right |^2 \bigl\langle\alpha_1 | \tilde\rho^i |
\alpha_1\bigr\rangle \ .
\label{fivesix}
\end{equation}
To compute the minimum in (\ref{fivetwo}) that defines ${\cal
S}(\tilde\rho^f, \tilde\rho_i)$, we should
choose the bases $\{|\alpha_k\rangle\}$ so as to satisfy
(\ref{fivefive}) and minimize
\begin{equation}
s(p) \equiv - \sum\nolimits_{\alpha_n, \cdots, \alpha_1} p (\alpha_n,
\cdots, \alpha_1) \log p(\alpha_n, \cdots, \alpha_1) \ .
\label{fiveseven}
\end{equation}
Less distributed sets of probabilities have smaller values of
$s(p)$. More precisely, consider two probability distributions
$p(\alpha_n, \cdots, \alpha_1)$ and $p^\prime(\alpha_n, \cdots,
\alpha_1)$ which differ only in that the probability of $\alpha_k$ is
distributed in the first, but exactly correlated with, say,
$\alpha_{k-1}$ in the latter.  That is, consider
\begin{equation}
p^\prime(\alpha_n, \cdots, \alpha_1) =
\delta_{\alpha_k\alpha_{k-1}}\sum\nolimits_{\alpha_k} p (\alpha_n,
\cdots, \alpha_1)\ .
\label{fiveeight}
\end{equation}
Then
\begin{equation}
s(p) - s(p^\prime) = -\sum\nolimits_{\alpha_n, \cdots, \alpha_1} p
(\alpha_n, \cdots, \alpha_1) \log \left[{p(\alpha_n, \cdots, \alpha_1)
\over \sum\nolimits_{\alpha_k} p(\alpha_n, \cdots,
\alpha_1)}\right]\ .
\label{fivenine}
\end{equation}
Since the $p$'s are positive numbers, this shows that
\begin{equation}
s(p^\prime) \leq s(p)\ .
\label{fiveten}
\end{equation}
Thus, we reduce $s(p)$ computed from the probabilities (\ref{fivesix})
 by aligning the bases as
much as possible so that they are exactly correlated from one time to
the next.  At the minimum the intermediate bases coincide with either
$\{|\alpha_1\rangle\}$ or $\{|\alpha_n\rangle\}$. This yields a
decoherent set if $\tilde\rho^i$ is diagonal in $\{|\alpha_1\rangle\}$
and $\tilde\rho^f$ is diagonal in $\{|\alpha_n\rangle\}$.

The sum $-\Sigma p(\alpha) \log p(\alpha)$ is
the same for a set of histories and a finer-grained set with
alternatives {\it exactly} correlated with those of the first and this is
true for an arbitrary number of times $n$. Thus, for the purposes of
computing ${\cal S} (\tilde\rho^f, \tilde\rho^i)$, completely
fine-grained histories may be replaced by histories of the form
\begin{equation}
C_\alpha = P^f_{\alpha_f} P^i_{\alpha_i}
\label{fiveeleven}
\end{equation}
where $P^f_{\alpha_f}$ are projections onto a basis
$\{|\alpha_f\rangle\}$ diagonalizing $\tilde\rho^f$ and $P^i_{\alpha_i}$
are projections onto a basis $\{|\alpha_i\rangle\}$ diagonalizing
$\tilde\rho^i$.  Such sets of histories are exactly decoherent.  A
simple expression for the probabilities of the histories
(\ref{fiveeleven}) can be found by summing
(\ref{fivethree}) over $\alpha^\prime$. Then
\begin{mathletters}
\begin{eqnarray}
p(\alpha_f, \alpha_i) &=& {\cal N}\ Tr\left(\tilde\rho^f P^f_{\alpha_f}
P^i_{\alpha_i} \tilde\rho^i\right) \\
                      &=& \frac{\tilde\pi^f_{\alpha_f}
\tilde\pi^i_{\alpha_i} |
\langle\alpha_f|\alpha_i\rangle|^2}{\sum\nolimits_{\alpha_f\alpha_i}
\tilde\pi^f_{\alpha_f} \tilde\pi^i_{\alpha_i}
|\langle\alpha_f|\alpha_i\rangle|^2}
\end{eqnarray}
\label{fivetwelve}
\end{mathletters}
where $\tilde\pi^f_{\alpha_f}$ and
$\tilde\pi^i_{\alpha_i}$ are the eigenvalues of $\tilde\rho^f$ and
$\tilde\rho^i$ respectively.  Thus
\begin{equation}
{\cal S} (\tilde\rho^f, \tilde\rho^i) =
\mathop{\min}_{\{|\alpha_f\rangle\}\ \{|\alpha_i\rangle\}}
\left[-\sum\nolimits_{\alpha_f \alpha_i} p(\alpha_f, \alpha_i) \log
p(\alpha_f, \alpha_i)\right]
\label{fivethirteen}
\end{equation}
where the minimum is taken over bases $\{|\alpha_f\rangle\}$ that
diagonalize $\tilde\rho^f$ and bases $\{|\alpha_i\rangle\}$ that
diagonalize $\tilde\rho^i$. Such a minimum is
still necessary for the definition because there will be
several different bases that diagonalize
$\tilde\rho^f$ and/or $\tilde\rho^i$ if the
$\{\tilde\pi^f_{\alpha_i}\}$ or $\{\tilde\pi^i_{\alpha_i}\}$ are
degenerate.  We shall illustrate in what
follows.

Eq (\ref{fivethirteen}) gives an explicit form for ${\cal S}(\tilde\rho^f,
\tilde\rho^i)$ in terms of the eigenvalues of $\tilde\rho^f$ and
$\tilde\rho^i$ and the bases that diagonalize them.  It is thus
completely determined by $\tilde\rho^f$ and $\tilde\rho^i$. We can
illustrate the construction with two special cases:

We first consider the case $\tilde\rho^f=I$ and $\tilde\rho^i\equiv
\tilde\rho$, a density matrix.  This is the case of a final condition of
indifference with respect to final state (which is no condition at all)
and an initial density matrix.
It coincides with usual quantum mechanics of Section II.
Any basis will diagonalize
$\tilde\rho^f = I$ and $\tilde\pi^f_{\alpha_f}=1$.
{}From (\ref{fivetwelve}) we then have
\begin{equation}
p(\alpha_f, \alpha_i) = \tilde\pi_{\alpha_i}
|\langle\alpha_f|\alpha_i\rangle|^2 \equiv \tilde\pi_{\alpha_i}
q^{\alpha_i}_{\alpha_f}
\label{fivefourteen}
\end{equation}
where $\tilde\pi_{\alpha_i}$ are the probabilities which are the
diagonal elements of the density matrix $\tilde\rho$.  Note that for
each $\alpha_i$, the $q^{\alpha_i}_{\alpha_f}$ are themselves a set of
probabilities, and we can write
\begin{equation}
-\sum\nolimits_{\alpha_f\alpha_i} p\left(\alpha_f, \alpha_i\right) \log
p\left(\alpha_f, \alpha_i\right) = s \left(\tilde\pi\right) +
\sum\nolimits_{\alpha_i} \tilde\pi_{\alpha_i} s
\left(q^{\alpha_i}\right) \ .
\label{fivefifteen}
\end{equation}
To find ${\cal S} (I, \tilde\rho)$ we minimize (\ref{fivethirteen}) over
{\it all} bases
$\{|\alpha_f\rangle\}$. Since $s(q^{\alpha_i})\geq 0$, the minimum is
obtained by choosing
$\{|\alpha_f\rangle\}$ to coincide with $\{|\alpha_i\rangle\}$.
All the
probabilities $q^{\alpha_i}_{\alpha_f}$ are then
either zero or one and
$s(q^{\alpha_i})=0$ for each $\alpha_i$. Thus,
\begin{equation}
{\cal S} \left(I, \tilde\rho\right) = -\sum\nolimits_{\alpha_i}
\tilde\pi_{\alpha_i} \log \tilde\pi_{\alpha_i} =
-Tr\left(\tilde\rho\log\tilde\rho\right)
\label{fivesixteen}
\end{equation}
\noindent In this way we derive the usual information measure on single density
matrices as the least missing information in fine-grained decoherent
sets of histories.

We should point out, however, that the limit $\tilde\rho^f\to I$ is not
smooth.  Consider $\tilde\rho^f=I+\epsilon B$ where $\epsilon$ is a
small parameter and $B$ is a Hermitian operator with non-degenerate
eigenvalues
in a basis $\{|\beta\rangle\}$. Then,
following through the above calculation, we find in the limit
$\epsilon\to 0$ that
\begin{equation}
\lim\limits_{\epsilon\to 0} {\cal S} \left(I+\epsilon B,
\tilde\rho\right) = -
Tr\left(\tilde\rho\log\tilde\rho\right)
+ \sum\nolimits_{\alpha_i} \tilde\pi_{\alpha_i} s (q^{\alpha_i})
\label{fiveseventeen}
\end{equation}
where the
$q^{\alpha_i}_\beta = |\langle \beta |\alpha_i\rangle |^2$ are now fixed.  The
limit
of ${\cal S}(\tilde\rho^f, \tilde\rho^i)$ as $\tilde\rho^f\to I$ therefore
depends
on the direction that $I$ is approached in the space of operators
$\tilde\rho^f$. The least of these limits is (\ref{fivesixteen}). The
largest might be as large as
 $-Tr(\tilde\rho\log \tilde\rho) + \log N$, where $N$ is
the dimension of the Hilbert space since $s(p) \leq \log N$. Similar
statements will apply in approaching any $\tilde\rho^f$ or
$\tilde\rho^i$ with degenerate eigenvalues.

The origin of this direction dependence may be intuitively understood as
follows:  ${\cal S}(\tilde\rho^f, \tilde\rho^i)$, as defined by
({\ref{fivethirteen}), measures not only how distributed the
probabilities $\tilde\pi^i_{\alpha_i}$ of the initial state are,
 and how distributed the
probabilities $\tilde\pi^f_{\alpha_f}$ of the final state are, but also how
distributed the probabilities are of the final states given an initial
state --- the quantities $|\langle\alpha_f|\alpha_i\rangle|^2$. As long
as $\epsilon$ is finite those quantities are fixed as $\epsilon\to 0$.
When $\epsilon$ is strictly 0, $\tilde\rho_f=I$ no longer singles out any
basis.  A
compression of information is possible from that needed to specify a
particular basis $\{|\alpha_f\rangle\}$ relative to the initial one to
the trivial statement that all bases are equivalent.

Another interesting case of the measure ${\cal S}(\tilde\rho^f,
\tilde\rho^i)$ occurs when $\tilde\rho^f$ and $\tilde\rho^i$ commute.  Then,
assuming no degeneracy, there is a unique {\it common} basis in which they are
diagonal and
\begin{equation}
p\left(\alpha_f, \alpha_i\right) = \delta_{\alpha_f\alpha_i}
\ \frac{\tilde\pi^f_{\alpha_f}
\tilde\pi^i_{\alpha_i}}{\sum\nolimits_\beta \tilde\pi^f_\beta
\tilde\pi^i_\beta}\ .
\label{fiveeighteen}
\end{equation}
(Cases where $\tilde\rho^f$ and/or $\tilde\rho^i$ are degenerate may be
discussed with arguments similar to those following (\ref{fiveseventeen}) with
the result that (\ref{fiveeighteen}) provides the minimizing
probabilities). The diagonal elements of (\ref{fivethirteen}) are those
of the density matrix
\begin{equation}
\bar\rho = \frac{\tilde\rho^f\tilde\rho^i}{Tr \left(\tilde\rho^f
\tilde\rho^i\right)}\ .
\label{fivenineteen}
\end{equation}
Thus, when $\tilde\rho^f$ and $\tilde\rho^i$ commute,
\begin{equation}
{\cal S}\left(\tilde\rho^f, \tilde\rho^i\right) = - Tr \left(\bar\rho
\log \bar\rho\right)\ .
\label{fivetwenty}
\end{equation}
This result might have been expected classically.  In classical physics,
where there is no non-commutation, there is a deterministic correlation
between initial and final conditions.  A restriction on histories
by a distribution of phase space initial conditions and a distribution
of final conditions is equivalent to a more restrictive distribution of initial
conditions and no restriction at all on final conditions.  That new initial
distribution is the product of the old one with the final condition
evolved back to the initial time.  In the Heisenberg picture we are
using, that product is the analog of (\ref{fivenineteen}) when
$\tilde\rho^f$ and $\tilde\rho^i$ commute.

The maximum possible value of ${\cal S}(\tilde\rho^f, \tilde\rho^i)$ is
attained when $\tilde\rho^f
= I + \epsilon B$ and $\tilde\rho^i = I + \epsilon E$ in the limit as
$\epsilon \to 0$ where $B$ and $E$ are operators whose diagonal
bases are maximally skewed $|\langle\alpha_f | \alpha_i\rangle|^2 =$
constant. Then, in the limit $\tilde\pi^f_{\alpha_i} = 1/N$ and
$\tilde\pi^i_{\alpha_i} = 1/N$ where $N$ is the dimension of the Hilbert
space, the probabilities $p\left(\alpha_f, \alpha_i\right) = 1/N^2$ are
as distributed as possible and
\begin{equation}
{\cal S}_{\rm max} = 2 \log N \ .
\label{fivetwentyone}
\end{equation}

The minimum possible value of ${\cal S}$ is zero since $-\sum p\, \log p$ is
positive or zero. Pairs of positive, Hermitian operators for which there is no
missing information are of special interest.  In order for ${\cal S}
(\tilde\rho^f, \tilde\rho^i)$ defined by (\ref{fivethirteen}) to vanish,
all but one of  the probabilities
$p(\alpha_f, \alpha_i)$ must vanish.  Call the labels of that
probability $(\hat\alpha_f,
\hat\alpha_i)$. Then, from (\ref{fivetwelve})
\begin{equation}
{\cal N} \tilde\pi^f_{\alpha_f} \tilde\pi^i_{\alpha_i} | \langle
\alpha_f | \alpha_i \rangle |^2 = \delta_{\alpha_f\hat\alpha_f}
\delta_{\alpha_i\hat\alpha_i}\ .
\label{fivetwentytwo}
\end{equation}
Two extreme cases illustrate some of the ways of satisfying the conditions
(\ref{fivetwentytwo}).  First suppose that $\langle \alpha_f |
\alpha_i\rangle \not= 0$ for all $(\alpha_f, \alpha_i)$.  Then
$\tilde\pi^f_{\alpha_f} = \delta_{\alpha_f\hat\alpha_f}$ and
$\tilde\pi^i_{\alpha_i} = \delta_{\alpha_i\hat\alpha_i}$. The
initial and final conditions are both pure states.  This is not a very
interesting case because the decoherence functional (\ref{fiveone})
factors and only trivial sets of histories can decohere.

At the opposite extreme the bases $\{|\alpha_i\rangle\}$ and
$\{|\alpha_f\rangle\}$ may coincide so that $\rho^f$ and $\rho^i$
commute. The condition (\ref{fivetwentytwo}) is then satisfied when
$\bar\rho$ given by (\ref{fiveeighteen}) is pure. This can happen
when either the initial or final state is pure. Thus, another
example is:
\begin{equation}
{\cal S} (\rho, |\psi\rangle\langle \psi |)=0\qquad {\rm if}
\qquad [\rho, |\psi\rangle\langle\psi|\,]=0\ .
\label{fivetwentythree}
\end{equation}
This includes the familiar case
\begin{equation}
{\cal S}(I, |\psi\rangle\langle \psi|)=0
\label{fivetwentyfour}
\end{equation}
that arises in usual quantum mechanics in (\ref{fivesixteen}).
There are many other ways of satisfying (\ref{fivetwentytwo}).

\subsection[]{The Missing Information in a Set of Histories}
\label{subsec:vc}

The information measure ${\cal S}(\tilde\rho^f, \tilde\rho^i)$ in pairs
of positive, Hermitian operators may now be used to define the missing
information in a decoherent set of histories $\{C_\alpha\}$ in
time-neutral generalized quantum mechanics in analogy
with the construction of Section II.  We assume we are given initial and
final operators $\rho^f$ and $\rho^i$ that define a decoherence
functional through (\ref{fiveone}). We define the missing information in
the set $\{C_\alpha\}$ by
\begin{equation}
S(\{C_\alpha\}) = \mathop{\max}_{\tilde\rho^f, \tilde\rho^i}
\quad \left[{\cal S} (\tilde\rho^f,
\tilde\rho^i)\right]_{\tilde D(\alpha^\prime, \alpha) = D(\alpha^\prime,
\alpha)}\ .
\label{fivetwentyfive}
\end{equation}
The maximum is taken over positive, Hermitian operators
$\tilde\rho^f$ and $\tilde\rho^i$ that preserve the value of the
decoherence functional for the histories $\{C_\alpha\}$ according to
\begin{equation}
\tilde D(\alpha^\prime, \alpha)=
\widetilde{\cal N}\ Tr\left(\tilde\rho^f C_{\alpha^\prime} \tilde\rho^i
C^\dagger_\alpha\right) = D(\alpha^\prime, \alpha) = {\cal N}\ Tr
\left(\rho^f C_{\alpha^\prime} \rho^i C^\dagger_\alpha \right)\ .
\label{fivetwentysix}
\end{equation}

If $\{\bar C_{\bar\alpha}\}$ is a coarse graining of $\{C_\alpha\}$ in
the sense of (\ref{twosixteen}), then the conditions for preserving the
decoherence functional of the coarser-grained set are
linear combinations of the conditions
 for preserving the finer-grained set.  Thus
\begin{equation}
S\left(\{\bar C_{\bar\alpha}\}\right) \geq S
\Bigl(\{C_\alpha\}\Bigr)\ .
\label{fivetwentyseven}
\end{equation}
Evidently $\tilde\rho^f = \rho^f$ and $\tilde\rho^i=\rho^i$ preserve the
decoherence functional so that
\begin{equation}
S(\{C_\alpha\}) \geq {\cal S} (\rho^f, \rho^i)\ .
\label{fivetwentyeight}
\end{equation}
The missing information in any set of histories is always greater than
the information measure of the operators defining the initial and final
conditions.

We define complete information about the system as the least missing
information in any set of decoherent histories
\begin{equation}
S_{\rm compl}= \mathop{\min}_{{\rm decoherent}\ \{C_\alpha\}}
S(\{C_\alpha\})\ .
\label{fivetwentynine}
\end{equation}
{}From (\ref{fivetwentyeight}) the minimum cannot be less than ${\cal
S}(\rho^f, \rho^i)$, and so
\begin{equation}
S_{\rm compl}\geq {\cal S}(\rho^f, \rho^i) \  .
\label{fivethirty}
\end{equation}
For generic initial and final operators $\rho^f$ and $\rho^i$ whose
non-zero eigenvalues are non-degenerate, we expect
(\ref{fivethirty}) to be an equality because preserving the
decoherence functional for the fine-grained set $\{C_\alpha=P^f_{\alpha_f}
P^i_{\alpha_i}\}$, where $\{P^f_{\alpha_f}\}$ and $\{P^i_{\alpha_i}\}$ are
projections onto bases, in (\ref{fivetwentysix}) uniquely determines
$\tilde\rho^f$ and $\tilde\rho^i$ up to trivial rescalings. However, for
those situations where the condition does not determine $\tilde\rho^f$
and $\tilde\rho^i$ uniquely, the maximum in (\ref{fivetwentyfive}) may be
larger than the minimum in (\ref{fivetwentynine}), and (\ref{fivethirty})
be only an inequality.  We illustrate with an example:

Consider a set
of histories $\{C_\alpha\}$ and
the case of usual quantum mechanics with $\rho^f = I$ in an
$N$-dimensional Hilbert space with $N$ even.
In absence of further argument we have no
reason to suppose that the $(\tilde\rho^f, \tilde\rho^i)$ that provide
the maximum in (\ref{fivetwentyfive}) are of the form $(I, \tilde\rho^i)$,
but suppose that to be the case. Then ${\cal S}$
is given by (\ref{fiveseventeen}) with the choice of the arbitrary basis
$\{|\beta\rangle\}$ such as to {\it maximize} its value.  The
maximum values of $s(q^{\alpha_i})$ are each $\log N$ and this can be
realized since, when $N$ is even,
 there is a unitary matrix all of whose elements have the
same absolute value.  Under these assumptions
\begin{equation}
S(\{C_\alpha\}) = S_{\rm usual}(\{C_\alpha\}) + \log N
\label{fivethirtyone}
\end{equation}
where $S_{\rm usual} (\{C_\alpha\})$ is the missing information in
$\{C_\alpha\}$ calculated according to the rules of usual quantum
mechanics as in Section II. Were (\ref{fivethirtyone}) to hold for every
set of decohering histories, it would follow that the complete
information would be
\begin{equation}
S_{\rm compl} = - Tr(\rho^i \log \rho^i) + \log N\ .
\label{fivethirtytwo}
\end{equation}
This is larger than ${\cal S}(I, \rho^i)$ by the addition of $\log N$
[{\it c.f.} (\ref{fivesixteen})], so that (\ref{fivethirty}) is only
an inequality.

Eq (\ref{fivethirtytwo}) is enough to show that even when $\rho^f=I$ the
notion of missing information in time-neutral quantum mechanics does not
necessarily coincide with that of usual quantum mechanics.  That is
because time-neutral quantum mechanics utilizes a notion  of information
that involves both initial and final conditions and the relation between
them. Loosely speaking there is more information to be missing.
However, were the two notions compared when $\rho^f = I$ the {\it
difference} in information between sets of histories
would be the same in both
formulations and that is what is needed to discriminate between sets
of histories by
their information content.

\subsection[]{Missing Information on Spacelike Surfaces}
\label{subsec:vd}

The time-neutral generalized quantum mechanics we have been discussing
cannot generally be reformulated in terms of states on spacelike surfaces.
However, we can discuss the information available in alternatives on a
spacelike surface and whether that information is the same on one
spacelike surface as on another.

Consider for simplicity a spacelike surface of constant time $t$ in a
particular Lorentz frame.  Alternatives at that moment of time are
represented by sets of orthogonal Heisenberg picture projection
operators $\{P_\alpha(t)\}$.  We define the missing information on the
surface of constant $t$ by
\begin{equation}
S(t) \equiv \mathop{\min}_{{\rm decoherent}\ \{P_\alpha(t)\}}
S(\{P_\alpha(t)\})\ .
\label{fivethirtythree}
\end{equation}
That is, the missing information at $t$ is the least of that missing in
all the alternatives at that time.

Whatever its value, $S(t)$ as defined
by (\ref{fivethirtythree})  is conserved.
That is because, in the Heisenberg picture,
 any set of projections $\{P_\alpha\}$ may be regarded as
projections on {\it some} quantity at {\it any} time.  Thus the same sets of
projection operators are available on any surface.  The minimum is
therefore the same on all surfaces:
\begin{equation}
S(t^\prime) = S(t^{\prime\prime})\ .
\label{fivethirtyfour}
\end{equation}

Is complete information available on any
surface?  That is the question of whether $S(t)$ defined by
(\ref{fivethirtythree}) coincides with $S_{\rm compl}$ as defined by
(\ref{fivetwentynine}).

Consider for simplicity the case when neither $\rho^i$ nor $\rho^f$ have
degenerate, non-zero eigenvalues and do not commute. Then $S_{\rm compl}$
is ${\cal S}(\rho^f, \rho^i)$. Complete information would be available
on any spacelike surface if there were a set of projections
$\{P_\alpha\}$ defining decoherent alternatives such that
$S(\{P_\alpha\})= {\cal S}(\rho^f, \rho^i)$. However, ${\cal S}(\rho^f,
\rho^i)$ is realized by fine-grained histories that are sequences of at
least {\it two} sets of projections [{\it cf.} (\ref{fiveeleven})].  One might
choose $\{P_\alpha\}$ to coincide with one or the other of these but not
both.  Therefore, generically $S(t)$ is greater than $S_{\rm compl}$.
Complete information about the initial and final conditions is not
available on any one spacelike surface in time-neutral quantum
mechanics.  Complete information {\it is} available through histories
that involve alternatives on at least two spacelike surfaces, although
these may be separated by only  an infinitesimal time.

\section[]{Information in Generalized Quantum Theories}
\label{sec:VI}

The discussion of spacetime information for
time-neutral generalized quantum mechanics in
the preceding section suggests how measures of information could be
constructed in arbitrary generalized quantum theories.  We describe that
construction in this section.

A generalized quantum theory of a closed system consists of three elements
\cite{Har91b}: (1) the sets of fine-grained histories which are the most
refined possible description of the system; (2) the allowed
coarse-grained sets of alternative histories, which generally are
partitions of some fine-grained set into mutually exclusive classes
$\{c_\alpha\}$; and (3) a decoherence functional
$D(\alpha^\prime, \alpha)$ measuring interference between pairs of
histories in a coarse-grained set.  The decoherence functional
is a complex
valued functional on pairs of histories which is~~(i) Hermitian, (ii)
positive, (iii) normalized, and (iv) consistent with the principle of
superposition in the specific senses described in \cite{Har91b}.  Given
these three elements a set of coarse-grained alternative histories
(approximately, medium) decoheres when the ``off-diagonal'' elements of
$D(\alpha^\prime, \alpha)$ are sufficiently small.  The diagonal
elements give the probabilities of the individual probabilities of the
individual histories in the decoherent
 set.  The rules both for which sets of histories
may be assigned probabilities and for the values of those
probabilities are thus summarized by the fundamental formula:
\begin{equation}
D(\alpha^\prime, \alpha) \approx \delta_{\alpha^\prime\alpha}
p(\alpha)\ .
\label{sixone}
\end{equation}

Two examples of generalized quantum theories have been described in
previous sections. The first is usual quantum theory.
Its fine-grained histories are defined by sequences of sets of
one-dimensional projections, one set at each time, with individual
histories represented by (continuous) chains of projections, one from each
set.  Coarse-grained histories are represented by class operators which
are sums of these, as in (\ref{twofour}). The decoherence functional is
given by (\ref{twofive}).  The time-neutral generalized quantum theory
of Section V is a second example.  The fine-grained and coarse-grained
sets of histories are the same as in usual quantum mechanics, but the
decoherence functional (\ref{fiveone}) is different, incorporating
both an initial and final condition.  Other examples are the generalized
quantum field theory in fixed background spacetimes with closed timelike
curves to be discussed in the next section and the generalized quantum
mechanics of dynamical spacetime geometry described in \cite{Harpp}.

What all these examples have in common is a decoherence functional
constructed from certain elements which represent histories and their
evolution and other elements which represent the quantum boundary
conditions.  Examples of the former are the projections, Hamiltonian
action, etc.~in the examples discussed.  Examples of the elements
specifying boundary conditions are the positive matrices $\rho^i$ and
$\rho^f$ representing initial and final conditions in (\ref{fiveone}).
Assuming such a division of the elements entering the decoherence
functional we can construct information measures as follows:

We first define the information content of the boundary conditions.
One way to do this would be to simply choose an information measure on
the elements of the decoherence functional that define the boundary
conditions.  We did this in Section II when we chose the measure ${\cal
S}(\tilde\rho)=-Tr(\tilde\rho\log\tilde\rho)$ for density matrices.
However, a more satisfactory approach is to define the measure, {\it
intrinsically}, in terms of the probabilities of a standard class,
${\cal C}_{\rm stand}$, of very fine-grained, decoherent, sets of
histories that probe these boundary conditions.  Specifically, we
define the missing
information in the boundary conditions, $S(D)$, as the least missing
 information in the probabilities of
the sets of histories $\{c_\alpha\}$ in the class ${\cal C}_{\rm stand}$
\begin{equation}
{\cal S}(D) = \mathop{\min}_{\{c_\alpha\}\in\ {\cal C}_{\rm
stand}}
\left[-\sum\nolimits_\alpha
p(\alpha)\log p(\alpha)\right]\ .
\label{sixtwo}
\end{equation}

A natural choice for the class ${\cal C}_{\rm stand}$ is the class of
decoherent
completely fine-grained sets of histories.  This was used to define the measure
${\cal S}(\tilde\rho^i, \tilde\rho^f)$ for the time neutral quantum
mechanics discussed in Section V.  However, for some generalized quantum
theories there may be no completely fine-grained sets that decohere.
For instance, this is likely to be the case in any sum-over-histories
generalized
quantum theory.  Another possibility for ${\cal C}_{\rm stand}$ would be
the class of fin{\it est}-grained sets of histories that decohere that is, the
class of sets which decohere but for which no finer graining decoheres.
However, this is a more difficult class to compute and it is not even
clear that the this choice would lead to the standard measure
$-Tr(\tilde\rho\log\tilde\rho)$ in usual quantum mechanics.

Once class ${\cal C}_{\rm stand}$ is chosen, or the measure ${\cal S}(D)$
otherwise fixed, the Jaynes construction can be
implemented
to define the missing information
in a set of coarse-grained decoherent histories $\{c_\alpha\}$.  The
missing information is the maximum of the information content of those
boundary conditions which reproduce the decoherence and probabilities of
the set $\{c_\alpha\}$.  That is,
\begin{equation}
S(\{c_\alpha\}) = \mathop{\rm max}\limits_{\tilde D} \left[{\cal S}
(\tilde D)\right]_{\tilde D(\alpha^\prime, \alpha) = D(\alpha^\prime,
\alpha)}
\label{sixthree}
\end{equation}
where the maximum is over all boundary conditions.  Evidently
\begin{equation}
S(\{c_\alpha\}) \geq {\cal S}(D)\ .
\label{sixfour}
\end{equation}

The missing information in a class ${\cal C}$ of sets of histories is
then straightforwardly defined as
\begin{equation}
S({\cal C}) = \mathop{\min}_{{\rm
decoherent}\atop \{c_\alpha\}\in {\cal C}} S(\{c_\alpha\})\ .
\label{sixfive}
\end{equation}
The least missing information in the class of {\it all} histories defines the
complete information:
\begin{equation}
S_{\rm compl}= \mathop{\min}_{{\rm decoherent}\ \{c_\alpha\}}
S(\{c_\alpha\})\ .
\label{sixfivea}
\end{equation}
Clearly,
\begin{equation}
S(\{c_\alpha\}) \geq S_{\rm compl} \geq {\cal S}(D) \ .
\label{sixfiveb}
\end{equation}
If there is at least one set of histories for which the condition
${\tilde D}(\alpha', \alpha)= D(\alpha',\alpha)$ implies that ${\tilde D}=D$,
then $S_{\rm compl}$ equals ${\cal S}(D)$. That was the case for
the usual quantum mechanics of Section II, but not the case for the
time-neutral quantum mechanics of Section V.

It is easily seen that the definitions (\ref{sixtwo}), (\ref{sixthree}),
and (\ref{sixfive}) coincide with the specific examples discussed in
previous sections, but provide a general and abstract framework which we
shall illustrate with another example in the next section.

\section[]{Fixed Spacetimes with Non-Chronal Regions}
\label{sec:VII}

Spacetime must be foliable by spacelike surfaces for the quantum
mechanics of matter fields to be formulated in terms of the unitary
evolution and reduction of a state vector defined on such surfaces.
However, not all spacetimes permit a foliation by spacelike surfaces.
Examples are spacetimes with closed timelike curves, such as might be
produced by the relative motions of wormhole mouths \cite{WORM}. For
such spacetimes a more general formulation of quantum mechanics is
required and a
number have been discussed \cite{Deu91},  \cite{KTpp},
\cite{FPS91}, \cite{Yur93}, \cite{Hawxx},  \cite{Har91a}, \cite{Har94},
\cite{And94}.
In this section we apply the general notions of information described in
the previous sections to generalized quantum theories suitable for
spacetimes with such non-chronal regions.

The generalized quantum theories we shall discuss were described in
\cite{Har94} and \cite{And94} whose notation we follow.  We briefly
review them here. We consider spacetimes with a fixed background
geometry having a compact non-chronal
region $NC$. We consider an initial region before $NC$ that is foliable
by spacelike surfaces and a final region after $NC$ also foliable by
spacelike surfaces.  The region in between, however, is not so foliable.
We consider the
quantum mechanics of a single scalar field $\phi(x)$ moving in this
background geometry.  The
characteristic feature of the theories we consider is that the
transition matrix between a state of definite spatial field
configuration $\phi^\prime(\bf x)$ on a spacelike surface
$\sigma^\prime$ before $NC$ and a similar state of definite spatial field
configuration $\phi^{\prime\prime}(\bf x)$ after $NC$ is generally
non-unitary. Transition matrices may be defined by sums-over-field histories
between $\sigma^\prime$ and $\sigma^{\prime\prime}$
\begin{equation}
\left\langle\phi^{\prime\prime}(\bf x), \sigma^{\prime\prime}|\phi^\prime
(\bf x), \sigma^\prime\right\rangle = \int\nolimits_{[\phi^\prime,
\phi^{\prime\prime}]} \delta\phi\,
\exp\bigl(iS[\phi (x)]\bigr)
\label{sevenone}
\end{equation}
where $S[\phi(x)]$ is the action functional for the
scalar field.  As suggested by Klinkhammer and Thorne \cite{KTpp}, and
demonstrated by Friedman, Papastamatiou, and Simon
\cite{FPS91}, the transition matrix defined by (\ref{sevenone}) is
non-unitary for an interacting field theory, order by order in
perturbation theory.

We can construct generalized quantum theories incorporating such non-unitary
evolution as follows: For the set of fine-grained histories we take
sequences of sets of one-dimensional projections on every member of a
foliating set of spacelike surfaces before $NC$ and every member of a
foliating family of spacelike surfaces after $NC$.  Simple examples of
coarse-grained sets may be represented in a Heisenberg-like picture by
chains of projections before $NC$
\begin{mathletters}
\begin{equation}
C_\alpha = P^k_{\alpha_k} (\sigma_k) \cdots P^1_{\alpha_1}(\sigma_1)\ ,
\quad \sigma_i< \sigma_-\ ,
\end{equation}
and chains of projections after $NC$
\begin{equation}
C_\beta = P^n_{\beta_n} (\sigma_n) \cdots P^{k+1}_{\beta_{k+1}}
(\sigma_{k+1}), \quad \sigma_i > \sigma_+ \ .
\end{equation}
\label{seventwo}
\end{mathletters}

\noindent where $\sigma_-$ is a spacelike
 surface just before $NC$ and $\sigma_+$ is a spacelike
surface just after $NC$. The projections in (\ref{seventwo}) evolve
unitarily both before and after $NC$.  A
non-unitary operator $X$, derived from (\ref{sevenone}), connects the
alternatives before $NC$ to those after $NC$ so that a whole history
consisting of sets of alternatives at sequences of times is represented
by
\begin{equation}
C_\beta X C_\alpha
\label{seventhree}
\end{equation}
with $C_\alpha$ and $C_\beta$ as given by (\ref{seventwo}).  More
general coarse grainings are obtained by partitioning such histories
into mutually exclusive classes with class operators which are the
corresponding sums of those of (\ref{seventhree}).

Two different forms of the decoherence functional give two distinct
generalized quantum theories incorporating non-unitary evolution.
The original proposal of \cite{Har94} was
to take, for histories of the form (\ref{seventhree}),
\begin{equation}
D\left(\beta^\prime, \alpha^\prime; \beta, \alpha\right) =
Tr\left[C_{\beta^\prime} X C_{\alpha^\prime} \rho C^\dagger_\alpha
X^\dagger C^\dagger_\beta\right]/Tr \left(X\rho X^\dagger\right)
\label{sevenfour}
\end{equation}
where $\rho$ is a density matrix representing the initial condition of
the closed system.  An attractive alternative, proposed by Anderson
\cite{And94}, is to take
\begin{equation}
D\left(\beta^\prime, \alpha^\prime; \beta, \alpha\right) =
Tr\left[X^{-1} C_{\beta^\prime} X C_{\alpha^\prime} \rho
C^\dagger_\alpha X^\dagger C_\beta (X^\dagger)^{-1}\right]
\label{sevenfive}
\end{equation}
assuming $X$ is invertible.  Both (\ref{sevenfour}) and
(\ref{sevenfive}) are easily seen to satisfy the general
requirements (i)--(iv) for decoherence functionals
mentioned in Section VI. The definition (\ref{sevenfive}) has the
advantages that, in contrast to (\ref{sevenfour}), it is
linear in the initial $\rho$ and does not lead to acausal effects in
which the non-chronal regions in the future can effect the probabilities
of present alternatives.  Which, if any, form emerges from a more
fundamental quantum theory of spacetime is an open question.  In this
section we shall concentrate on implementing the notions of information
described in Section VI using Anderson's (\ref{sevenfive}). A similar
but not coincident discussion could be given on the basis of
(\ref{sevenfour}).

A word is in order concerning the relation of the generalized quantum
theories under discussion to the sum-over-field-histories formulations
described in \cite{Har94}. In a sum-over-histories formulation the set
of fine-grained histories are the possible four-dimensional, spacetime
field configurations, $\phi(x)$, and coarse grainings are restricted to
partitions of these into mutually exclusive classes.  Partitions by the
values of spatial field configuration $\phi({\bf x})$ on a spacelike
surface outside of $NC$
 defines one kind of coarse graining of field histories whose class operators
can be
represented by projection operators as in (\ref{seventwo}).
However, partitions of the fields can also be used to define
alternatives {\it inside} $NC$, for instance
a partition by ranges of values of the
field averaged over a spacetime region inside $NC$.  The class operators
of such spacetime alternatives, defined by functional integrals over the
appropriate class of fields, are not generally projection operators or
even of the form (\ref{seventhree}). The generalized quantum theories we
are discussing in this section are both more restricted and more general
than such sum-over-histories formulations.  They are more restricted
because they do not deal with alternatives inside $NC$ but only on
spacelike surfaces outside of $NC$.  On the other
hand, outside $NC$, the alternatives are more general.  The alternatives
of a sum-over-field-histories formulation would be restricted to projections
onto ranges of spatial field configurations.  By contrast, the present
discussion considers  all the alternatives available by transformation
theory provided they are defined on spacelike surfaces outside $NC$.  We
consider information of this particular example of generalized quantum
theory, not because it is more general or
more fundamental than the sum-over-field-histories formulation,
but because its closer
connection with the usual quantum theory discussed in Section II
makes it a more useful example.

In the
language of Section VI, $\rho$ is the element in the decoherence
functionals (\ref{sevenfour}) and (\ref{sevenfive}) representing
the boundary condition, and the $P$'s, $X$, $H$, etc.~are the
elements
representing the histories and their evolution.
Our first task, therefore, is to find the
measure of missing information in a density matrix $\tilde\rho$ in the
presence of the non-unitary evolution $X$. This is defined by
(\ref{sixtwo}) and denoted by ${\cal S}_X(\tilde\rho)$.  We take the
class of all fine-grained sets of histories of the form (\ref{seventwo})
for the class ${\cal C}_{\rm stand}$. Then, specifically,
\begin{equation}
{\cal S}_X(\tilde\rho) = \mathop{\min}_{{\rm
fine-grained}\atop {\rm decoherent}\ \{C_\beta, C_\alpha\}}
\left[-\sum\nolimits_{\alpha\beta} \tilde p(\beta, \alpha) \log \tilde p
(\beta, \alpha)\right]\ .
\label{sevensix}
\end{equation}
The fine-grained sets of histories are chains of sets of
one-dimensional projections, one on each spacelike surface outside $NC$.
Extending the analysis of Section V one easily sees that for the purpose
of computing ${\cal S}_X(\rho)$ one may consider histories of the form
\begin{equation}
C_\alpha = P^i_\alpha = |\alpha\rangle\langle\alpha|\ ;
\quad C_\beta =
P^f_\beta = |\beta\rangle\langle\beta|
\label{sevenseven}
\end{equation}
where $\{|\alpha\rangle\}$ is a basis in which
$\rho$ is diagonal and $\{|\beta\rangle\}$ is a basis which $XX^\dagger$ is
diagonal.  Decoherence requires all other finer-grained alternatives to
be exactly correlated with these.

The probabilities of the decoherent set (\ref{sevenseven}) are
\begin{equation}
\tilde p(\beta, \alpha) = \tilde\pi_\alpha | \langle\beta | X |
\alpha\rangle |^2 \langle\beta | (XX^\dagger)^{-1} |\beta\rangle
\equiv \tilde\pi_\alpha q^\alpha_\beta\ .
\label{seveneight}
\end{equation}
It is easily seen that, for fixed $\alpha$, the numbers $q^\alpha_\beta$
are probabilities and that, similarly to (\ref{fivefifteen}),
\begin{equation}
-\sum\nolimits_{\alpha\beta} \tilde p (\beta, \alpha) \log \tilde p
(\beta, \alpha) = s (\tilde\pi) + \sum\nolimits_\alpha \tilde \pi_\alpha
s(q^\alpha) \ .
\label{sevennine}
\end{equation}
Thus, we have for the information measure of the initial condition in
$\tilde\rho$ the presence of a non-unitary $X$.
\begin{equation}
{\cal S}_X (\tilde \rho) = -Tr(\tilde\rho \log \tilde \rho) +
\mathop{\rm min}\limits_{\{|\beta\rangle\}} \sum\nolimits_\alpha
\tilde\pi_\alpha s(q^\alpha)
\label{seventen}
\end{equation}
where the minimum is over bases that diagonalize $XX^\dagger$. If there
is a unique basis that diagonalizes $XX^\dagger$, then (\ref{seventen})
gives an explicit formula for ${\cal S}_X(\tilde\rho)$.

Were $X$ unitary, we could pick the orthogonal basis $\{|\beta\rangle\}$
to be $\{X^\dagger |\alpha\rangle\}$ thereby making $q^\alpha_\beta =
\delta^\alpha_\beta$ and ${\cal S}_X (\tilde\rho) = {\cal S} (\tilde
\rho)$.  However, when $X$ is non-unitary we have only
\begin{equation}
{\cal S}_X (\tilde\rho) \geq {\cal S} (\tilde\rho) \equiv - Tr
\left(\tilde\rho \log \tilde\rho \right)\ .
\label{seveneleven}
\end{equation}

\noindent Eq (\ref{seveneleven}) shows that ${\cal S}_X(\tilde\rho)$ generally
does not
coincide with ${\cal S}(\tilde\rho)$. In the presence of a
domain non-unitary evolution somewhere in the spacetime, the missing
information in a density matrix is greater than it would be if the
domain had not been present.  That is because the missing information in
$\tilde\rho$ has been defined in terms of the probabilities of the
finest-grained decoherent histories which it predicts, and those
histories extend over the whole of spacetime --- both before and after
any non-chronal region.

With the definition of the information content of an initial $\rho$ in
hand, the missing information in a set of histories $\{C_\beta,
C_\alpha\}$ of the form (\ref{seventhree}) can be straightforwardly defined
from the general schema
(\ref{sixthree})
\begin{equation}
S_X\left(\{C_\beta\}, \{C_\alpha\}\right) \equiv \mathop{\rm
max}\limits_{\tilde\rho} \left[{\cal S}_X(\tilde\rho)\right]_{\tilde
D(\alpha^\prime, \alpha) = D(\alpha^\prime, \alpha)}\ .
\label{seventwelve}
\end{equation}
Missing information in a class of histories is defined by
(\ref{sixfive}) and complete information, $S_{\rm compl}$, by the minimum of
(\ref{seventwelve}) over all decoherent sets of histories. Evidently,
$S_{\rm compl} \geq {\cal S}_X(\rho)$, but we shall show in the following
that
\begin{equation}
S_{\rm compl}= {\cal S}_X(\rho)\ ,  \label{seventwelvea}
\end{equation}
by exhibiting one example for which the equality is satisfied.

This generalized quantum mechanics of fields in non-chronal backgrounds
cannot be reformulated in terms of states on a spacelike surface, their
unitarily evolution between such surfaces, and their reduction at them.
However, we may still investigate how much information about the
system is available in histories that are confined to a spacelike surface
$\sigma$.
Clearly there are two types of surfaces --- those before the non-chronal
region $NC$ and those after it.  We define
\begin{mathletters}
\begin{equation}
S_{\rm before}(\sigma) = \mathop{\rm min}\limits_{{\rm decoherent}
\ \{P_\alpha(\sigma)\}} S_X\left(I, \{P_\alpha(\sigma)\}\right)\ ,
\label{mlett:thirteen a}
\end{equation}
for $\sigma \leq \sigma_-$, and similarly, for $\sigma \geq \sigma_+$,
\begin{equation}
S_{\rm after}(\sigma) = \mathop{\rm min}\limits_{{\rm decoherent}
\ \{P_\beta(\sigma)\}} S_X\left(\{P_\beta(\sigma)\}, I\right)\ .
\label{mlett:thirteen b}
\end{equation}
\label{seventhirteen}
\end{mathletters}
In the Heisenberg picture, a set of projection operators
$\{P_\alpha\}$ is a projection onto ranges of the values of {\it some}
quantity on {\it any} surface.  Therefore, the minimum (\ref{mlett:thirteen a})
will be the same on all surfaces $\sigma$ before $NC$.  Similarly for
(\ref{mlett:thirteen b}) after $NC$.  Thus, missing information $S_{\rm
before}(\sigma)$ is conserved before $NC$ and missing information
$S_{\rm after}(\sigma)$ is conserved after $NC$. It is not immediately
obvious, however, whether information is conserved in passing from
before to after $NC$.  We now show that it is for a reasonably generic
set of cases, and that complete information is available on each
spacelike surface outside $NC$.

We consider the case where $XX^\dagger$ is independent of $X\rho
X^\dagger$ in a sense made precise below.
For surfaces before $NC$ consider the missing information in a set of
projections $P_\alpha = |\alpha\rangle\langle \alpha|$ onto a basis that
diagonalizes $\rho$.  According to (\ref{seventwelve}) this is the
maximum of ${\cal S}_X(\tilde\rho)$ over all $\tilde\rho$ which
reproduce the decoherence functional for this set of projections.  The
condition that the decoherence functional be reproduced is
\begin{eqnarray}
\tilde D(\alpha^\prime, \alpha) &=& Tr\left[X^{-1} X P_{\alpha^\prime}
\tilde\rho P_\alpha X^\dagger (X^{-1})^\dagger\right]\nonumber \\
&=& \left\langle\alpha^\prime |\tilde\rho|\alpha\right\rangle =
D(\alpha^\prime, \alpha) =
\left\langle\alpha^\prime|\rho|\alpha\right\rangle\ .
\label{sevenfourteen}
\end{eqnarray}
Thus, $\tilde\rho=\rho$ is the unique density matrix which reproduces
the decoherence functional.  From (\ref{seventwelve})
\begin{equation}
S_X\bigl(I, |\alpha\rangle\langle\alpha|\bigr) = {\cal S}_X(\rho)=S_{\rm
compl}\ .
\label{sevenfifteen}
\end{equation}
This one example is enough to demonstrate the equality in
(\ref{seventwelvea}).
Thus, complete information is available on every spacelike surface
before $NC$.

For surfaces after $NC$ consider sets of projections $P_\beta =
|\beta\rangle\langle\beta|$ onto a basis which diagonalizes the
Hermitian operator $X\rho X^\dagger$. This set is decoherent.  The
condition that a density matrix $\tilde\rho$ reproduce the decoherence
functional for this set is
\begin{eqnarray}
\tilde D(\beta^\prime, \beta) &=& \langle\beta |(X X^\dagger)^{-1}
|\beta^\prime\rangle \langle\beta^\prime | X\tilde\rho
X^\dagger|\beta\rangle\nonumber \\
&=& D(\beta^\prime, \beta) = \delta_{\beta\beta^\prime} \langle\beta
|(X X^\dagger)^{-1}| \beta\rangle \langle\beta | X\rho
X^\dagger | \beta\rangle\ .
\label{sevensixteen}
\end{eqnarray}
If we assume that $\langle\beta | (X
X^\dagger)^{-1}|\beta^\prime\rangle$ has no non-vanishing matrix
elements, then we can conclude that $X\tilde\rho X^\dagger = X\rho
X^\dagger$. Since $X$ is assumed invertible we have $\tilde\rho=\rho$
for the unique $\tilde\rho$ that reproduces the decoherence functional
of $\rho$ for the set $\{P_\beta\}$.  Thus, from (\ref{seventwelve})
\begin{equation}
S_X\bigl(|\beta\rangle \langle\beta |, I) = {\cal S}_X(\rho\bigr)=S_{\rm
compl}
\  .
\label{sevenseventeen}
\end{equation}
Thus, complete information is available about the quantum system on
every spacelike surface after $NC$.

Taken together, (\ref{sevenfifteen}) and (\ref{sevenseventeen}) show that,
despite the absence of states on a spacelike surface,
complete information about the quantum system is
available on every spacelike surface outside of $NC$.  Complete
information is conserved; it is the same ${\cal S}_X(\rho)$ on all
spacelike surfaces, both before and after $NC$. However,
the conserved,
complete, missing information ${\cal S}_X(\rho)$ on any surface in a
spacetime with a non-chronal region is greater
than the missing information ${\cal S}(\rho) = -Tr(\rho \log \rho)$ in a
spacetime without such a non-chronal region.  That is the case even
though the predictions of Anderson's generalized
quantum mechanics coincide exactly with the usual
theory for spacelike surfaces before $NC$.  The reason is that the
missing information ${\cal S}_X(\tilde\rho)$ has been defined in terms
the probabilities of fine-grained sets of decoherent histories of the
closed system.  These extend arbitrarily far into future and
thus are affected by the existence of any non-chronal
region. The probabilities of these finest-grained decoherent histories
are more distributed because of the non-unitary evolution arising from
the non-chronal region than they would be without [{\it cf.}
(\ref{seveneight})]. Thus, the missing information ${\cal S}_X(\rho)$ is
greater than ${\cal S}(\rho)$.  One might be tempted to define
missing information in $\rho$
before $NC$ by the usual $-Tr(\rho \log \rho)$ since the predictions of
this generalized quantum mechanics coincide with the usual theory there.
That, however, would lead to an unexplained loss of information in
passing from before the non-chronal region to after it.  Here, we have
consistently adopted a {\it spacetime} approach to information with the
result that information
is conserved in passing from one spacelike surface to
another.

\section[]{Black Holes}
\label{sec:VIII}

Hawking's 1974 \cite{Haw74} prediction of a steady flux of thermal
radiation in test fields from black hole background spacetimes raised the
possibility that black holes could evaporate completely.  As a consequence,
information as usually defined in terms of states on spacelike surfaces,
would be permanently lost in the process of evaporation.  The questions
of the outcome of the Hawking process and its consistency with the basic
principles of quantum mechanics has been of intense interest since.  In
this section we discuss these
questions utilizing the notions of spacetime information developed in this
paper in the context of a generalized quantum theory of spacetime
geometry.

As yet we have no theory of quantum gravity adequate for
predicting the history of an evaporating black hole when it has shrunk
to less than Planck scale dimensions, despite a number of interesting
models \cite{snowup}. No improvement on this situation is offered there.
Rather, we describe information in the black hole evaporation process {\it
assuming}
that the black hole evaporates completely.  We do this in the
kinematical framework for a generalized quantum mechanics developed in
\cite{Harpp}. The qualitative information theoretic issues we
shall discuss are probably insensitive to the details in this framework
but it provides a reasonably concrete, if formal, setting in which to
consider them. We briefly recall some of its relevant features:

As mentioned in Section VI, there are
three elements in a generalized quantum theory: (1) the fine-grained
histories, (2) the allowed coarse-grained sets of histories,
and (3) a decoherence
functional defining the notion of interference between pairs of
histories in coarse-grained sets. For a theory of black hole
evaporation, we take the
fine-grained histories (1) to be a class of asymptotically flat spacetime
geometries with matter fields whose Penrose diagrams have the form shown
in Figure 1a.  We are thus dealing with a sum-over-histories generalized
quantum theory.  The precise nature of the geometries interior to the
asymptotic region -- how differentiable they
are, and what kinds of singularities are permitted, {\it etc} --  are
central issues in the specification of a complete theory of quantum
spacetime.  We do not resolve these issues here, but we assume that the
class of fine-grained histories at least includes those (Figure 2b)
whose Penrose diagram are of the kind
commonly taken to describe the complete evaporation of a black hole.
The coarse-grained histories (2) are partitions
of these fine-grained histories into mutually, exclusive,
 diffeomorphism-invariant classes
$\{c_\alpha\}$. In calculating the analogs of transition amplitudes, for
example, we are typically interested in partitions of the histories by
invariant descriptions of their asymptotic geometries and matter fields
on ${\cal I}^-, I^-$ and ${\cal I}^+, I^+$, leaving the interior fields
and geometry unrestricted.  The remaining element (3) is the decoherence
functional. This is specified first by constructing amplitudes for the
individual histories $c_\alpha$ in a coarse-grained set by sums over the
corresponding class of histories of the schematic form
\begin{equation}
\int\nolimits_{c_\alpha}\delta g\delta\phi \exp (iS[g, \phi])
\label{eightone}
\end{equation}
where $S[g, \phi]$ is the action of gravity coupled to matter fields
$\phi(x)$
and the integral is over geometries and fields in the class $c_\alpha$
with additional restrictions necessary to incorporate the boundary
conditions.  The decoherence functional $D(\alpha^\prime, \alpha)$ is a
bilinear
combination of these amplitudes analogous to (\ref{fiveone}).

We have deliberately been brief in sketching the details of such a
putative quantum kinematics of spacetime geometry because we wish to
make only one point: Such a generalized quantum mechanics of spacetime
geometry is not formulated in terms of states on spacelike surfaces or
their unitary evolution between such surfaces.  Neither is it likely
that it {\it can} be so formulated since the fine-grained histories single out
no set of spacelike surfaces to supply the preferred time usually
required in such theories. The notions of information, and of complete
information in particular, must be reexamined in such a theory.  The
discussion in Section VI provides a general framework for doing so.

The information measure of the boundary conditions, ${\cal S}(D)$, is the
least missing information in the
probabilities of the sets of histories is the class ${\cal C}_{\rm
stand}$ which, for definiteness, we may take to be the class of
 finest-grained decoherent sets of  histories.
(The argument is not strongly dependent on the choice for ${\cal C}_{\rm
stand}$.) Coarser-grained sets of
 histories will generally have more missing information.
Complete information is the least missing information among all sets of
decohering histories. Complete information is available in some set of
histories, but
complete information may not be available from histories
defined by alternatives on a single spacelike surface.  Further, the
most nearly complete information available on one spacelike surface is not
guaranteed to be the same as that available on another spacelike
surface.

Suppose the fine-grained histories describing a theory of transitions
between asymptotically flat regions contain those
 with Penrose diagrams as in
Figure 2b commonly assumed to describe the complete evaporation of a
black hole.  Then it seems especially likely that the same information
is not available on the surface
${\cal I}^+$ as it was on ${\cal I}^-$.  Plausibly some
information has gone down the black hole.

When quantum mechanics is in spacetime form and information formulated
in terms of spacetime histories complete information may not be
available on any particular spacelike surface.  Rather,
it may be necessary to search about the
four-dimensional spacetime to find complete information.
Histories defined by partitions near any horizon as well as partitions
near infinity may be necessary.  Thus, the absence of complete
information on a spacelike surface after the complete evaporation of a
black hole is not a violation of the principles of quantum mechanics
suitably generally stated.
Rather it becomes an interesting example of
the utility of formulating the theory in fully spacetime form.

\section[]{Conclusions}
\label{sec:IX}

Complete descriptions of quantum systems in terms of a state on a
spacelike surface can be expected only in situations when there is an
unambiguous notion of ``spacelike surface'' and the notion of causality
implicit in such a state apply.  In more general cases, where spacetime
geometry fluctuates, as in a quantum theory of gravity, or where it is
fixed, but lacking a foliating family of spacelike surfaces, as in
spacetimes with non-chronal regions, or when the
boundary conditions are inconsistent with usual causality, as in the
time-neutral formulation, we cannot
expect to formulate quantum mechanics in terms of states on spacelike
surfaces.  Rather a more general formulation of quantum
mechanics is needed.  In these circumstances it seems natural to
formulate quantum mechanics in fully spacetime form both with respect to
dynamics and alternatives. For such generalizations, the usual quantum
mechanical notions of information must also be generalized.  This paper
has provided one such generalization.  We have provided a general schema
for defining the complete information content in the boundary conditions
of a generalized quantum theory.  We have defined the information
available about these boundary conditions in a set of alternative
histories of the closed system and in classes of such sets.  These
notions of information are in fully spacetime form.  Complete
information may not be available on any spacelike surface.  Rather it
may be distributed about spacetime and available only through histories
that are not specific to any one spacelike surface.

\acknowledgments

The author is grateful for many conversations with Murray Gell-Mann on
the subjects of entropy and information in quantum mechanics over a long
period of time.  Thanks are due to A.~Connes, R.~Griffiths, J.~Halliwell,
C.~Isham, and N.~Linden, and S.~Lloyd
 for useful remarks and criticisms.  The author is
grateful to the Isaac Newton Institute and the Aspen Center for Physics
for hospitality while this work
was finished.  It was supported in part by the NSF grant PHY90-08502.

\figure{\bf Fig 1}:~~{Three examples of spacetime regions.
The region $R_a$ contains
a Cauchy surface $\sigma$.  Complete information is therefore available
in $R_a$.  Parts of a wave packet moving away from region $R_b$ will never
intersect it at any later time.  Complete information is therefore
unlikely to be available in $R_b$.  The region $R_c$ is the domain of
dependence of a region $L$ of a Cauchy surface $\sigma$.  The missing
information in $R_c$ is the same as the missing information in the
reduced density matrix for $L$, {\it i.e.}, the density matrix
$\rho(\sigma)$ traced over all field variables outside $L$.}

\figure{\bf Fig 2}:~~{In a generalized quantum mechanics describing black hole
evaporation the fine-grained histories are a class of
asymptotically flat geometries with Penrose diagrams of the form shown
in (a). If the class contains geometries with Penrose diagrams like that
of (b), that are usually said to describe the evaporation,  then it is
unlikely that the generalized quantum theory can recast as a theory of
evolving states on spacelike surfaces.  Complete information therefore
may not be available on any one spacelike surface and in particular not
on the  surface ${\cal I}^+$, because, plausibly information has gone
down the black hole.}

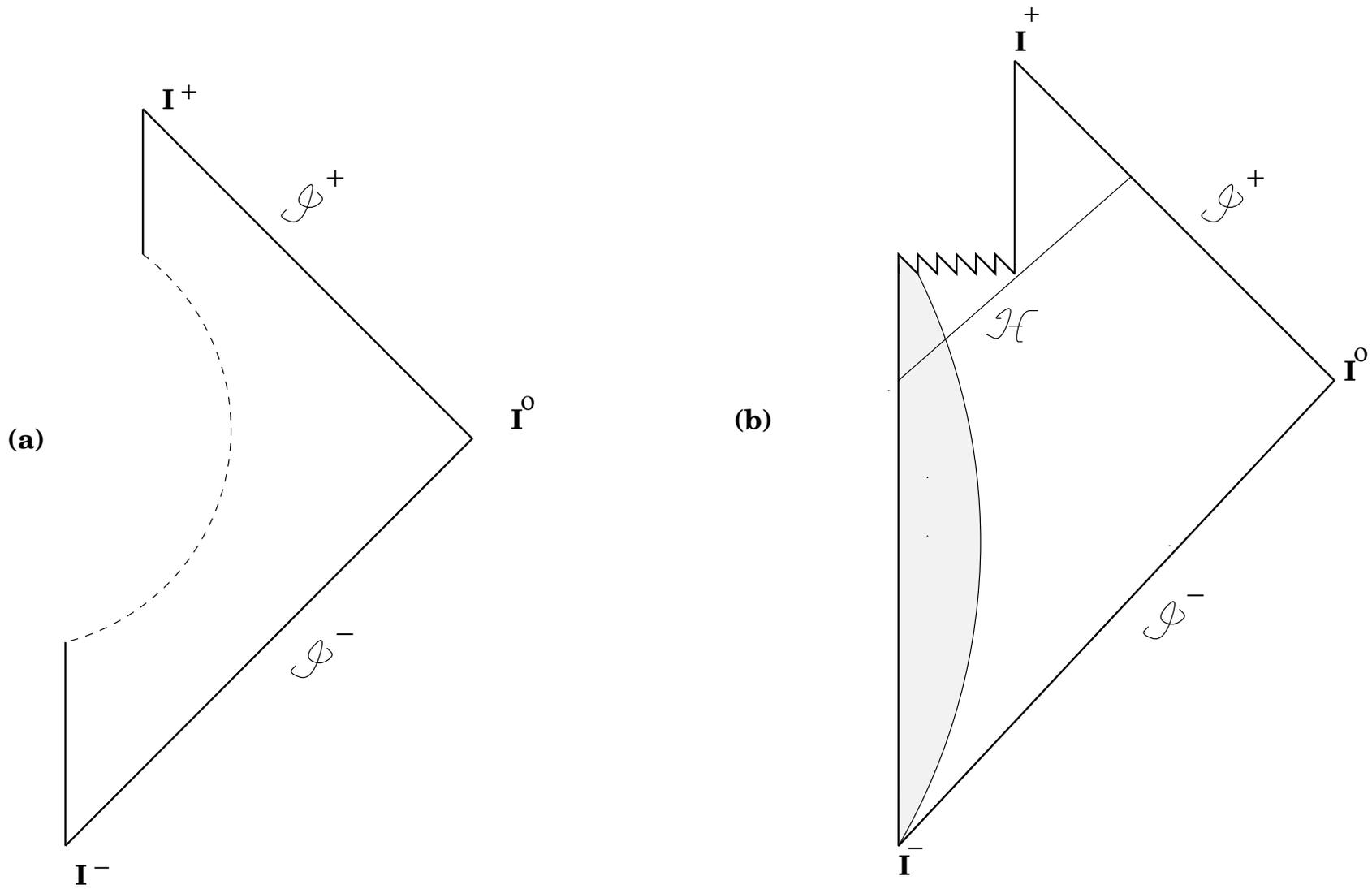
\end{document}